\documentclass[a4paper,superscriptaddress,twocolumn,apj,nofootinbib]{aastex631}
\usepackage{aas_macros}
\usepackage{acro}
\acsetup{patch/longtable=false}
\usepackage{tabularx}
\usepackage{graphicx}
\usepackage{dcolumn}
\usepackage{bm}
\usepackage{comment}
\usepackage{amsmath}
\usepackage{float}
\usepackage{lipsum}
\usepackage{color}
\usepackage{times}
\usepackage{textcomp}
\usepackage{latexsym}
\usepackage{mathtools}
\usepackage{url}
\usepackage[utf8]{inputenc}
\usepackage[normalem]{ulem}
\usepackage{physics}
\usepackage{subfigure}
\usepackage{soul}
\usepackage{hyperref}
\hypersetup{colorlinks,linkcolor={blue},citecolor={blue},urlcolor={red}}  
\setcitestyle{authoryear,open={},close={}}
\setcitestyle{open={(},close={)}}

\begin{document}
\title{Inferring properties of dark galactic halos using strongly lensed gravitational waves}

\author{Eungwang Seo}
\thanks{e.seo.1@research.gla.ac.uk}
\affiliation{SUPA, School of Physics and Astronomy, University of Glasgow, Glasgow G12 8QQ, United Kingdom}
\affiliation{Department of Physics, The Chinese University of Hong Kong, Shatin, NT, Hong Kong}

\author{Tjonnie G. F. Li}
\affiliation{Department of Physics, The Chinese University of Hong Kong, Shatin, NT, Hong Kong}
\affiliation{Institute for Theoretical Physics, KU Leuven, Celestijnenlaan 200D, B-3001 Leuven, Belgium}
\affiliation{Department of Electrical Engineering (ESAT), KU Leuven, Kasteelpark Arenberg 10, B-3001 Leuven, Belgium}
\author{Martin A. Hendry}
\affiliation{SUPA, School of Physics and Astronomy, University of Glasgow, Glasgow G12 8QQ, United Kingdom}

\begin{abstract}
Gravitational waves (GWs) can be deflected, similarly to electromagnetic (EM) waves, by massive objects through the phenomenon of {\em gravitational lensing\/}. 
The importance of gravitational lensing for GW astronomy is becoming increasingly apparent in the GW detection era, in which nearly 100 events have already been detected. 
As current ground-based interferometers reach their design sensitivities, it is anticipated that these detectors may observe a few GW signals that are strongly lensed by the dark halos of intervening galaxies or galaxy clusters.
Analysing strong lensing effects on GW signals is, thus, becoming important to understand the lens' properties and correctly infer the intrinsic GW source parameters.
However, one cannot accurately infer lens parameters for complex lens models with only GW observations because there are strong degeneracies between the parameters of lensed waveforms.
In this paper, we discuss how to conduct parameter estimation of strongly lensed GW signals and infer the lens parameters using additional EM information, including the lens galaxy's axis ratio and the GW source-hosting galaxy's lensed images.
We find that for simple spherically symmetric lens models, the lens parameters can be well recovered using only GW information.
On the other hand, recovering the lens parameters requires systems in which four or more GW images are detected with additional EM observations for non-axially symmetric lens models.
Combinations of GW and EM observations can further improve the inference of the lens parameters.
\end{abstract}

\section{Introduction}
Measuring the masses of the dark halos of galaxies or galaxy clusters is one of the most important challenges in astrophysics and cosmology. 
The halo masses are a key factor for understanding important phenomena relevant to the physics of galaxies, from galaxy formation and evolution~\citep{behroozi2010comprehensive, fanidakis2012evolution,wetzel2012galaxy} to galaxy quenching~\citep{woo2013dependence,zu2016mapping,wang2018elucid,behroozi2019universemachine}. 
Most methods developed to date for measuring masses of galaxies depend only on electromagnetic (EM) wave observations. 
Application of photometry to gravitational lensing events and spectroscopy for velocity dispersion measurements of galaxies and clusters have been widely used -- and have estimated the mass of galaxies with mild error ranges~\citep{broadhurst1994mapping,hoekstra2003well,mandelbaum2006galaxy,hansen2009galaxy,elahi2018using,kafle2018need}. 
However, EM waves can be easily absorbed and re-emitted by baryonic matter.
Also, accurate measurements of galaxy components' dynamics are necessary, which are challenging.

In this regard, one promising approach to measuring dark halo masses is to use gravitational waves (GWs), which are not subject to some of the same limitations as EM waves.
GWs can also be gravitationally lensed by massive objects like EM waves -- resulting in larger angular areas on the lens plane~\citep{wang1996gravitational,nakamura1998gravitational,takahashi2003wave,sathyaprakash2009physics}.
Unlike EM waves, the universe is nearly transparent to GWs, and lens mass measurements using GWs are, in principle, more straightforward by analysing their strain amplitudes. 

Among the different kinds of GW lensing that may be considered, strong lensing occurs when there are massive objects along the line of sight, such as galaxies and galaxy clusters, which have a larger Schwarzschild radius than the wavelengths of the propagating GW signals.
This results in a splitting of the original unlensed GW signal into multiple lensed signals, each characterised by a distinct amplitude, arrival time and phase shift.
Until now, confident evidence of strong lensing has not been found in the events detected during the first three observing runs (referred to as O1, O2 and O3, respectively) carried out by the LIGO, Virgo and KAGRA collaborations~\citep{hannuksela2019search,abbott2021search,abbott2023search}.
In the near future, however, it is expected that up to $\mathcal{O}(1 \sim 2)$ strongly lensed compact binary coalescences per year may be observed as ground-based GW detectors reach their design sensitivities~\citep{biesiada2014strong,ding2015strongly,ng2018precise,li2018gravitational,mukherjee2021inferring,wierda2021beyond}.

To investigate the properties of the lens from analysis of lensed GW signals, a lens model has to be specified, which characterises the properties of dark galactic halos.
In previous works, axially symmetric lens models were mainly considered for simplicity~\citep{cao2014gravitational,hannuksela2019search,hou2020gravitational,robertson2020does,abbott2021search,cheung2021stellar,seo2022improving}.
However, non-axially symmetric lens models should be considered for more realistic galactic scale lensing scenarios (e.g., \cite{kormann1994isothermal,golse2002pseudo,tessore2015elliptical}).

In this paper, we perform Bayesian parameter estimation and employ rejection sampling techniques to infer the properties of the dark halos of intervening galaxies using various lens models: 1) simple lens models that have previously been used, such as a point mass and singular isothermal sphere (SIS) and 2) more complicated lens models that can be non-isothermal and have ellipticity.
Furthermore, we also calculate the lensing cross-section of each density profile to investigate which lens model is more effective in generating GW-strong lensing.

To maintain consistency with prior analyses (e.g., \cite{abbott2021search,abbott2023search}), we assume the so-called 'Planck 15' cosmology, $H_0 = 67.8$ km $s^{-1} Mpc^{-1}$, $\Omega_{m}=0.308$,  $\Omega_{K}=0$ and $\Omega_{\Lambda}=0.692$~\citep{ade2016planck}.

\section{Strong lensing configuration}\label{sec:2}
\begin{figure}[ht]
\centering
\resizebox{7.5cm}{!}{\includegraphics{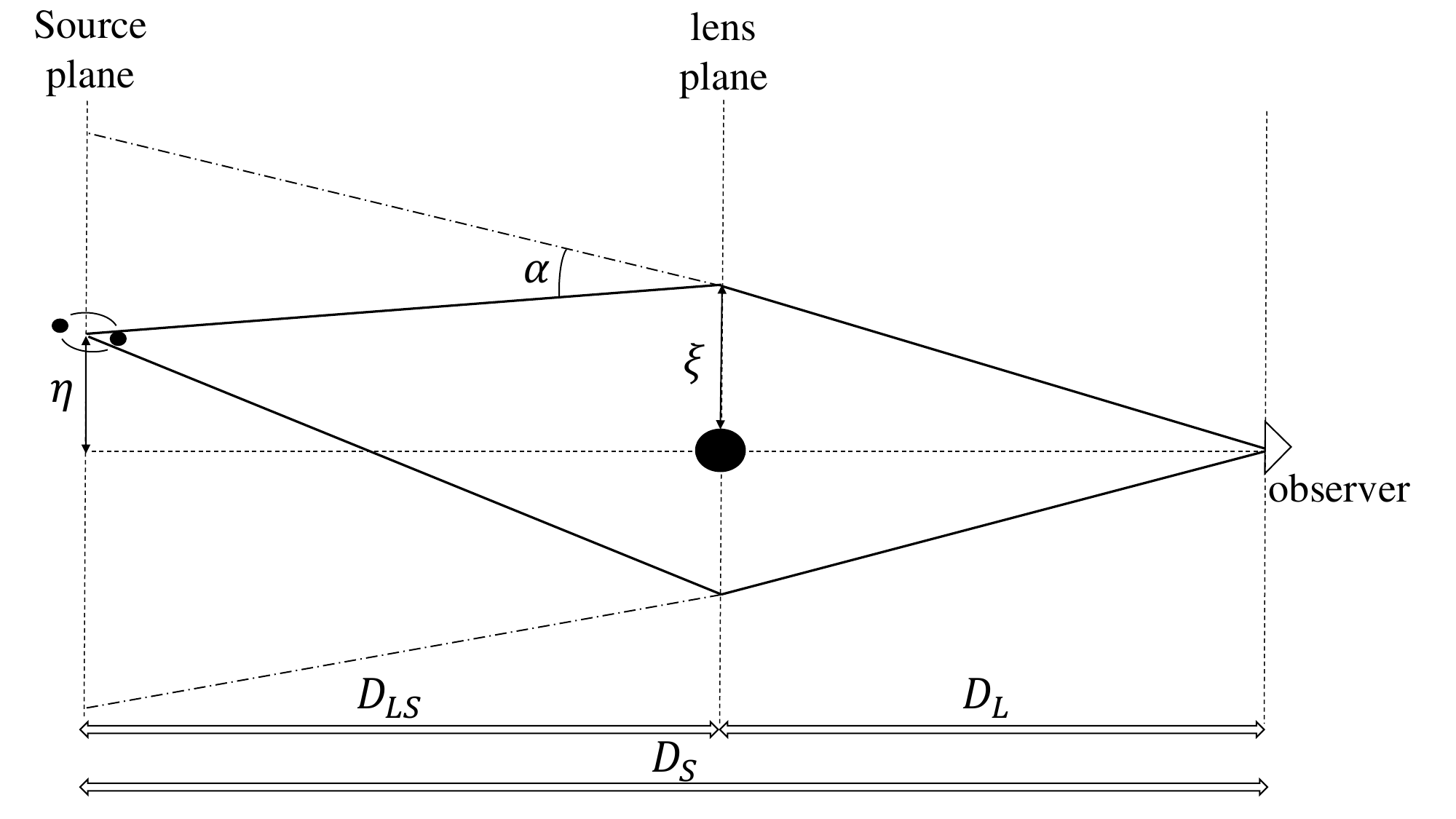}}\hspace{5pt}
\caption{Schematic of the gravitational lensing system. GW signals from a binary black hole on the source plane go through different paths to produce multiple images by a dark galactic halo on the lens plane. $\xi$ is the displacement between lens and image positions, the so-called impact parameter and $\eta$ is the source position from the line of sight.  $D_{L}$ and $D_{S}$ are the angular diameter distances between the observer and source and the observer and lens plane, respectively, and $D_{LS}$ is the angular diameter distance between the source and lens plane.} \label{schematic}
\end{figure}

The thin lens approximation holds in astrophysical gravitational lensing systems since distances between the observer and the source and lens are very large.
Thus, one can describe a gravitational lensing system using a two-dimensional plane.
Figure.~\ref{schematic} shows the geometry of the system in detail, where
$D_{L}$, $D_{S}$, and $D_{LS}$ are the angular diameter distances between observer and lens, observer and source, and source and lens, respectively.
It is convenient to express the image and source positions in dimensionless values.
\begin{equation}
    x=\frac{|\vec{\xi}|}{\xi_{0}}, \:\:
    y=\frac{D_{L}|\vec{\eta}|}{D_{S}\xi_{0}} ,
\label{position}
\end{equation}
where $\vec{\xi}$ and $\vec{\eta}$ are the displacements from the lens to an image and from the line of sight to the source, respectively and $\xi_{0}$ is a normalisation constant which makes $x$ and $y$ dimensionless.
In this work, we set the Einstein radius ($r_E$) as the normalisation constant if there is no specific indication otherwise.
The Einstein radius of an axially symmetric lens is defined by
\begin{equation}
\label{einstein_radius}
r_E=\sqrt{\frac{4 G M(r_E)}{c^2} \frac{D_L D_{L S}}{D_S}},
\end{equation}
where $M(r_{E})$ is the mass within the Einstein radius~\citep{schneider2006gravitational};
see Appendix for details of Einstein radii of various lens models.

Since we assume a dark galactic halo as our lens scale, which has a larger Schwarzschild radius than the wavelength of the propagating GWs from typical binary black hole systems detected by current ground-based interferometers, we can apply the geometrical optics approximation to calculate the lensed GW signal~\citep{wang1996gravitational,takahashi2003wave,dai2017waveforms}.
Within the geometrical optics regime, the lensed GW signal $h_{l}$ and unlensed GW signal $h_{u}$, both defined in the frequency domain $f$, have the following relationship,
\begin{equation}
\label{lensedwaveform}
h_{l}(f,\boldsymbol{\theta_{s}},\boldsymbol{\theta_{l}},M^{z}_{l})=F_{\rm geo}(f,\boldsymbol{\theta_{l}},M^{z}_{l})\times h_{u}(f,\boldsymbol{\theta_{s}}) ,
\end{equation}
where $\boldsymbol{\theta_{s}}$ and $\boldsymbol{\theta_{l}}$ are parameters of the GW source and model-dependent lens parameters respectively, and $M^{z}_{l} \equiv M_{l}(1+z_{l})$ is the redshifted lens mass of the intervening halo.
$F_{\rm geo}$ is the amplification factor, which is given by (see~\cite{takahashi2003wave})
\begin{equation}
\label{Fgeo}
F_{\rm geo}(f,\boldsymbol{\theta_{l}},M^{z}_{l}) = \sum_{j}|\mu_{j}|^{0.5}  \exp{i\pi\left(\frac{8fGM^{z}_{l}}{c^3} T_{j} - n_{j}\right)} ,
\end{equation}
where $\mu_{j}$ is the magnification factor of the $j$-th image, $x_{j}$, and $T_{j}$ and $n_{j}$ are the Fermat potential and Morse index of the $j$-th image, respectively.

Magnification factors and Fermat potentials can have different values for the same $\boldsymbol{\theta_{l}}$ depending on what lens model is assumed.
A lens model describes how much an image is distorted at an image position, with its deflection potential.
Thus, it is useful to show how deflection potentials determine the two parameters above.
The magnification factor and the Fermat potential for the $j$-th image can be written as
\begin{eqnarray}
\mu_j =\frac{1}{\operatorname{det}\left(\delta_{a b}-\frac{\partial^2 \psi_j}{\partial x_a \partial x_b}\right)}, \nonumber\\
T_j=\frac{1}{2}\left(|\vec{x}_{j}-\vec{y}|\right)^2-\psi_j+\phi_m,
\label{eq:mu_fp}
\end{eqnarray}
where $\psi_j$ is the deflection potential at the $j$-th image and $x_a$ indicates the $a$-component of $x_j$.
The $\phi_m$ term is the arrival time of the unlensed signal unchanged by lensing effects.

Using Eq.~\ref{eq:mu_fp}, one can also calculate the differences between ($\mu_{j}, T_{j}$) and ($\mu_{k}, T_{k}$), for the $j$-th image and $k$-th image, which we identify as the relative magnification factor $\mu_{\rm rel}$ and dimensionless time delay $\Delta T$.
\begin{equation}
\label{eq:murel}
\mu_{\rm rel} \equiv \frac{\mu_j}{\mu_k} =\frac{\operatorname{det}\left(\delta_{c d}-\frac{\partial^2 \psi_k}{\partial x_c \partial x_d}\right)}{\operatorname{det}\left(\delta_{a b}-\frac{\partial^2 \psi_j}{\partial x_a \partial x_b}\right)} .
\end{equation}
\begin{align}
\label{eq:dt}
\Delta T &\equiv \abs{T_k - T_j} \nonumber \\&= \abs{\frac{1}{2} \left(|x_k-y|^2 - |x_j-y|^2\right) + \psi_k -\psi_j}.
\end{align}
These two quantities, which we call \emph{lensing observables}, can be obtained numerically or analytically depending on the assumed lens model.
Furthermore, an actual time delay between the two images can be derived from Eq.~\ref{eq:dt} with a given redshifted lens mass under thin lens approximation~\citep{takahashi2003wave};
\begin{equation}
\label{mlzfromt}
    \Delta t = \frac{4GM^{z}_{l}}{c^3} \Delta T,
\end{equation}
where the $M^{z}_{l}$ is the redshifted mass within the normalisation factor mentioned in Eq.~\ref{position}.
Note that we do not deal with the Morse phase because it has just three discrete values $n_j=0/0.5/1$ when a Type I / Type II / Type III image occurs, respectively~\citep{dai2017waveforms}.
\section{Lens models}\label{lensmodel} 
In this section, we explore relatively simple (axially symmetric) and complex (non-axially symmetric) lens models, which are used to simulate various GW lensing systems on a galactic scale.
For some simple lens models, the lensing observables are analytically obtained from $\boldsymbol{\theta_{l}}$.
On the other hand, it is non-trivial to calculate the lensing observables from $\boldsymbol{\theta_{l}}$ for more realistic lens models.
One needs to solve the lens equations $y = x_{j}-\nabla \psi_{j}$ to obtain the $j$-th image positions and deflection potential values of each image in order to calculate the lensing observables.
\subsection{Axially symmetric lens}\label{spherelens}
We introduce three axially symmetric mass distributions as outlined in~\cite{takahashi2003wave}.
Firstly, the point mass (PM) model is one of the simplest density profiles to describe small objects, such as stars and dwarf star clusters whose Einstein radii are much larger than their physical sizes.
Under the geometrical optics approximation, lensing effects due to an isolated point mass are governed by the dimensionless source position  (i.e., $\boldsymbol{\theta_{l}}=y$) (See Eq.~\ref{position}), and it always generates two GW images.
Note that the PM model is unsuitable for describing galactic halos, but we consider the PM model to show the robustness of the inference technique presented in this paper.

Secondly, the singular isothermal sphere (SIS) is the simplest extended lens model used to characterize the distribution of matter in a gravitational system, such as a galaxy or a galaxy cluster. This model assumes that the velocity dispersion of particles in the system is constant at all radii.
Like the PM model, the value of $y$ determines the strong lensing effects due to SIS.
However, the SIS generates two images only when $y < 1$; otherwise, image splitting does not occur.
Lensing observables for the PM and SIS models can be analytically calculated as a function of $y$.

Lastly, the Navarro-Frenk-White (NFW) model is a density profile identified in numerical N-body simulations, assuming $\Lambda$CDM cosmology~\citep{navarro1997universal}.
Due to its dependence on various cosmological parameters, the strong lensing effects of the NFW model are more intricate to determine than those of PM and SIS models~\citep{wright2000gravitational,takahashi2003wave,oguri2019strong}.
Nevertheless, the dimensionless surface density ($\kappa_{s}$) includes information about the parameters, which is written as
\begin{equation}
\label{kappa}
\kappa_{s}=\frac{4\pi G}{c^{2}}\frac{D_{L}D_{LS}}{D_{S}}\rho_{s}r_{s}~,
\end{equation}
where $\rho_s$ and $r_s$ are the characteristic density and the scale radius of the lens~\citep{bartelmann1996arcs,keeton2001catalog}.
Thus, we define the lens parameters of the NFW model as $\boldsymbol{\theta_{l}}=(y,\kappa_{s})$.
Depending on $\boldsymbol{\theta_{l}}$, one or three images can be created with this model.
Also, unlike the PM and SIS models, the NFW model does not have an analytical form for the lensing observables. 
Therefore, one needs to numerically derive the lensing observables using Eqs.~\ref{eq:murel} and~\ref{eq:dt}, which requires solving the lens equations.
Details of the lensing formalism for these three axially symmetric lens models can be found in App.~\ref{appendix_simple}.
\subsection{Non-axially symmetric lens}\label{ellip_lens}
For more realistic scenarios, it is appropriate to use non-axially symmetric lens models with more lens parameters, such as the minor to major axis ratio, $q$.
Normally, unlike simple lenses, lensing observables of elliptical lens models should be numerically calculated from image positions and deflection angles.
However, the deflection angle depends on the position angle with respect to the centre of the halo as well as the displacement to the image position.
Thus, it is more demanding to solve the lens equations for non-axially symmetric lens models.

We introduce three elliptical mass distribution models generally used in EM strong lensing analyses~(e.g.,~\cite{treu2010strong, tessore2015elliptical,oguri2019strong,nightingale2019galaxy,shajib2021dark}).
First, the singular isothermal ellipsoid (SIE) is the simplest elliptical lens model~\citep{kormann1994isothermal}.
The SIE model assumes that the SIS model still approximately describes the mass distribution of the galaxy or cluster of galaxies but with some ellipticity.
Thus, the lensing effects of the SIE model are described by two parameters, $\boldsymbol{\theta_{l}}=(q,y)$.
Two or four images can be generated depending on the combination of the two lens parameters.

A more general lens model can no longer be isothermal.
The singular power-law ellipsoidal mass distribution (SPEMD) is one of the generally used elliptical and non-isothermal density profiles~\citep{barkana1998fast,tessore2015elliptical}, which has an additional lens parameter $\gamma$, the slope of halo density.
Hence, the lens parameters are $\boldsymbol{\theta_{l}}=(q,y,\gamma)$.
The SPEMD model can be reduced to an SIE when the density slope $\gamma=1$.
When $\gamma \geq 1$, two or four images can be generated like for the SIE, while three or five images can be generated when $\gamma < 1$.

Finally, we consider the elliptical Navarro-Frenk-White (ENFW) model~\citep{golse2002pseudo}, which is a modification of the standard spherical NFW profile described in Sec.~\ref{spherelens}.
Similar to the SIE and SPEMD cases, the strong lensing effects of the ENFW model are determined by $q$ and the lens parameters of the NFW ($\boldsymbol{\theta_{l}}=(q,y,\kappa_{s})$).
GW lensing systems described by the ENFW model can produce one, three or five images.
The details of these three non-axially symmetric lens models and their lensing formalism are presented in App.~\ref{appendix_complex}.

Given the lens model and lens parameters $\boldsymbol{\theta_{l}}$, one can solve the lens equations.
For each aforementioned lens model, we choose specific $\boldsymbol{\theta_{l}}$, corresponding to multiple solutions of the lens equations.
Subsequently, we calculate the corresponding relative magnification factors $\mu_{\rm rel}$ and dimensionless time delays $\Delta T$ for each $\boldsymbol{\theta_{l}}$.
These sets of parameters and lensing observables are then organised as ordered triplets ($\Theta = \{\boldsymbol{\theta_{l}}, \mu_{\rm rel}, \Delta T\}$) and stored in a \emph{lensing observable bank}.

\section{Strong lensing cross section}

It is informative to investigate how efficiently various lens models create multiple GW images when the same source redshift ($z_s$), lens redshift ($z_l$) and lens mass ($M_{l}$) are given.
The strong lensing cross-section ($\sigma$) can be defined as the area where multiple images form, magnified above a certain magnification threshold, $\mu_{\rm th}$ ~\citep{schneider2006gravitational,fedeli2006fast,robertson2020does}.
For a simple point-like lens, a disk with a radius equal to the Einstein radius ($r_{E}$) of the lens represents the strong lensing cross-section.
However, the strong lensing cross-sections of the complex lens models considered in Sec.~\ref{lensmodel} cannot be analytically calculated.

Since we can regard a GW source as a point source, the strong lensing cross-section is simply the area on the source plane enclosed by the caustics of the lens, where multiple images are created.
In this work, however, we impose an additional condition on the definition of the lensing cross-section for realistic detections, requiring that all images are magnified, i.e. $\mu_{j} > 1.0, \forall \, j$.
When a source is placed within an area $A_{s}$ satisfying the above conditions, the lensing cross-section can be written as
\begin{equation}
\sigma_{1.0}=\frac{\xi_{0}D_{S}}{D_{L}} \int_{A_{\mathrm{s}}} \mathrm{d}^2 y~,
\label{cross-section}
\end{equation}
where $y$ is the dimensionless unit on the source plane (See Eq.\ref{position})~\citep{fedeli2006fast}.

\begin{figure*}[t]
\centering
   \subfigure[\; $\sigma_{\rm 1.0}$ where at least \textit{two images} are created and magnified]
{\includegraphics[width=0.48\linewidth]{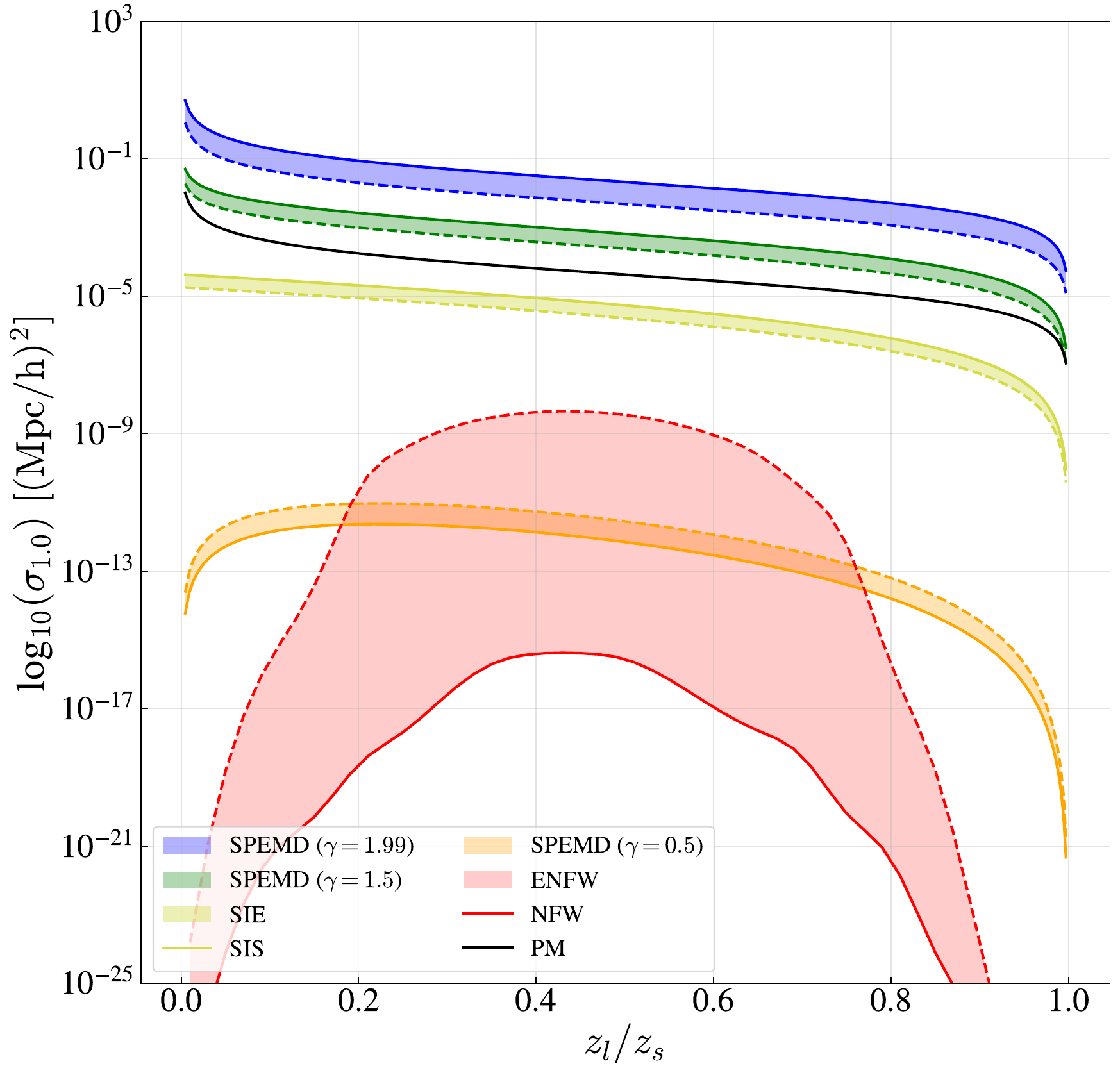} \label{lc_double}}
\subfigure[\; $\sigma_{\rm 1.0}$ where at least \textit{four images} are created and magnified]{\includegraphics[width=0.48\linewidth]{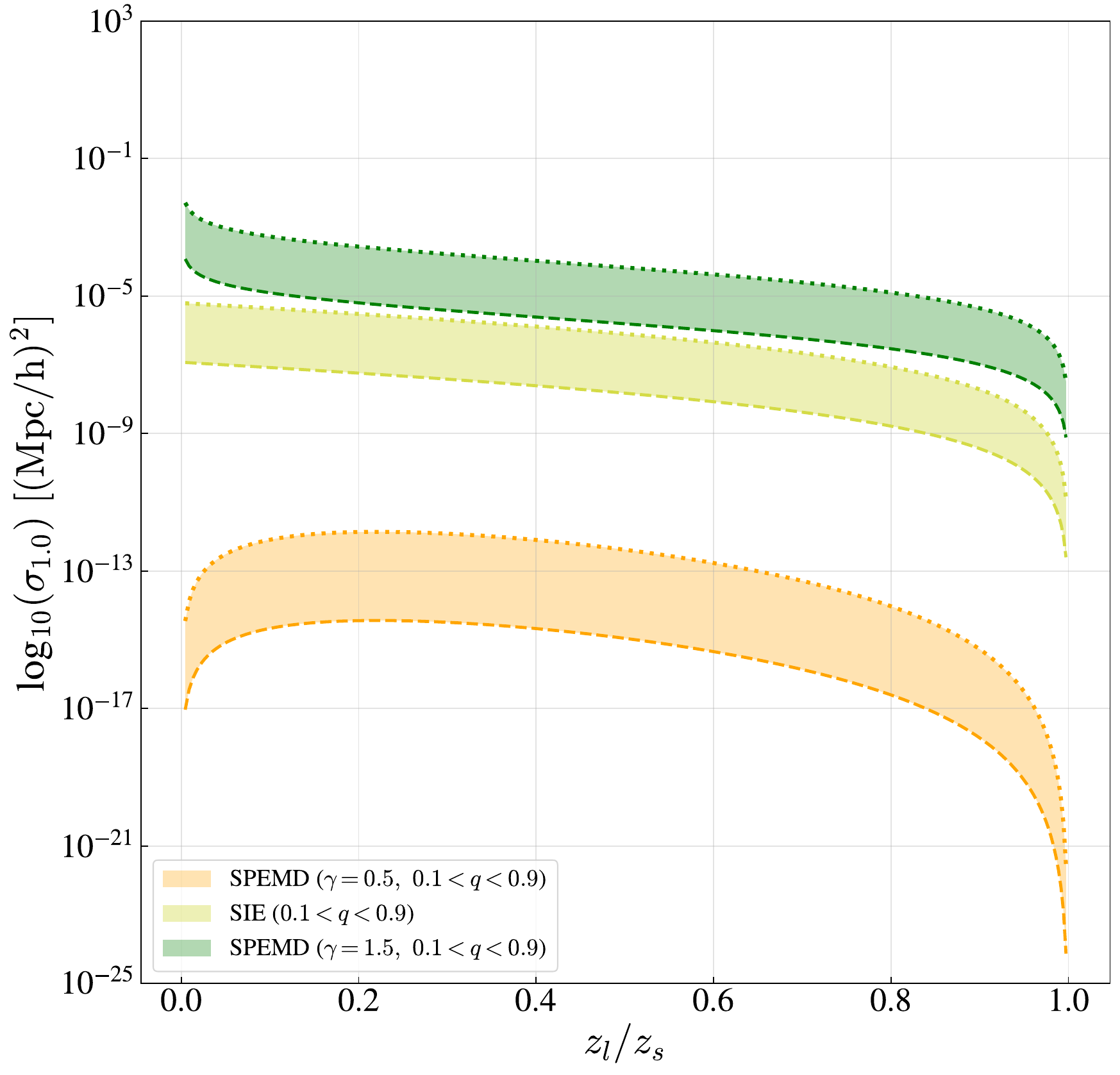}\label{lc_quadruple}}
\caption{\textit{Left panel}: Strong lensing cross section as a function of $z_{l}/z_{s}$, where at least two GW images are magnified (i.e., $\mu > 1.0$) for 
(E)NFW (red), SPEMD with $\gamma=0.5,~1.5,~1.99$ (orange, green, blue), SIE (yellow), and PM (black) lens models.
Coloured solid lines indicate the $\sigma_{\rm 1.0}$ of spherical lenses ($q=1.0$), and dashed lines do $\sigma_{\rm 1.0}$ of elliptical lenses with $q=0.1$.
Generally, it is more probable for denser lens models to create magnified images.
Furthermore, a SPEMD lens closer to the observer has larger $\sigma_{\rm 1.0}$, while the value of $\sigma_{1.0}$ for an (E)NFW lens decreases when the lens is closer to the observer because $\sigma_{1.0}$ depends on the critical density $\Sigma_{\rm cr}$.
Moreover, a lens with high ellipticity is less efficient in generating magnified images when $\gamma \geq 1$, while a lens with $\gamma < 1$ has the contrary aspect.
\textit{Right panel}: Same as the left panel, but at least four lensed GW signals are magnified.
Coloured dotted lines indicate $\sigma_{1.0}$ of elliptical lenses with $q=0.5$.
We set the maximum value of the axis ratio as $q=0.9$ because a lens with $q>0.9$ has a diminutive inner caustic area.
The function of $\sigma_{1.0}$ are similar to those of the double-image cases shown in Fig.~\ref{lc_double}, but the value of $\sigma_{1.0}$ changes more steeply with the ellipticity.
\label{fig:lc}}
\end{figure*}
We adopt the ray-shooting technique to calculate the area $A_s$~\citep{schneider1986two,kayser1986astrophysical}.
Firstly, we calculate the largest caustic (cut)\footnote{A cut arises when a lens model has a singularity at its centre.
When a source is located within the cut, multiple lensed images can be created even though the area is not enclosed by a caustic. The cut transforms into a caustic when a finite-size core is introduced at the centre of the lens.} area $B_s$ on the dimensionless source plane, given the lens models and their $\theta_{l}$ except for $y$.
Note that $y$ determines not the lens properties but the lensing effects on propagating GWs.
By using the ray-shooting technique, the corresponding image positions and magnification factors can be calculated for all source positions within $B_s$.
The ratio between the area filled with the source positions satisfying $\mu(y) > 1.0$ and the entire $B_s$ will be the dimensionless cross-section of interest, $A_s$.

Next, we adopt the Einstein radius $r_{E}$ (for PM and SPEMD-like models) and the scale radius $r_s$ (for E(NFW) models) as the normalisation constant $\xi_0$ that converts dimensionless values into real values.
The Einstein radius of a PM lens is a function of Einstein mass, $M_{E}$, but the Einstein radii of the SIS, SIE, and SPEMD lenses depend explicitly on the velocity dispersion, $\sigma_{v}$, of a galaxy rather than the $M_{E}$ of the galaxy.
For this reason, we assume a plausible galaxy that has an Einstein mass $M_{E} \sim 10^{12}M_{\odot}$ and velocity dispersion $\sigma_{v} \sim 200$~km/s, based on \cite{zahid2016scaling}.

For E(NFW) models, we assume another, different galactic halo whose mass within the scale radius is $M(r_{s})=10^{13}M_{\odot}$.
This is because the E(NFW) lenses with $M(r_{s})=10^{12}M_{\odot}$ create multiple images only vanishingly rarely due to their inefficiency at image splitting. 

The strong lensing cross-sections with at least two magnified images are shown for various lens models in Fig.~\ref{lc_double}.
We include the six lens models considered.
Among these lens models, elliptical models can have a double-image system for any value of $q$, but we consider only $q \geq 0.1$ for practicality.
For the SPEMD, we input three different density slope parameters ($\gamma=0.5,1.5,1.99$) to see how the density of a lens affects the lensing cross-section.

The SPEMD models, including SIS and SIE at a lower $z_l$ tend to have a higher $\sigma_{1.0}$ value.
Furthermore, $\sigma_{1.0}$ increases with the slope $\gamma$ of the density, which means that a denser lens model is more efficient at generating multiple signals. 
Note that the function of $\sigma_{1.0}$ of the SPEMD with $\gamma=1.99$ is very similar to that of PM because when $\gamma \rightarrow 2$, the SPEMD lens becomes a point mass~\citep{tessore2015elliptical}.
Moreover, a SPEMD lens with $\gamma < 1$ is more efficient in splitting and magnifying GW signals when the lens has high ellipticity (low $q$).
A SPEMD lens with $\gamma \geq 1$ and a higher ellipticity, on the other hand, has less efficiency in generating multiple magnified signals.

Non-axially symmetric lenses have multiple caustics, in which a propagating GW signal can be split into quadruple or more signals.
In this regard, we also calculate $\sigma_{1.0}$ for lens systems which can create quadruple or more lensed GW signals.

Fig.~\ref{lc_quadruple} shows strong lensing cross-sections $\sigma_{1.0}$, generating and magnifying at least four lensed GW signals.
The shapes of the plots are the same as the plots in Fig.~\ref{lc_double} since the unit areas of each lens model are the same as $\theta_{E}^{2}$.
However, the rate of change of $\sigma_{1.0}$ becomes more pronounced as the ellipticity varies.
Also, $\sigma_{1.0}$ has the maximum value at $q=0.5$, attributed to the limited region capable of producing magnified images when $q\rightarrow0.1$, and the shrinking of the caustic area as $q\rightarrow1$.
Note that we exclude the SPEMD with $\gamma=1.99$ and ENFW models as their inner caustics are too small to create quadruple or more images in a galaxy-scale gravitational lensing scenario.
\section{Inferring the parameters of a galaxy lens}
We simulate nine GW lensing systems based on the lens models described in the previous section.
For simplicity, we inject the same intrinsic unlensed signal $h_{\rm{u}}$ (GW150914-like event with luminosity distance $d_{L}=3~\rm{Gpc}$) and redshifted lens mass value $M^{z}_{l}=10^{12}\rm{M}_{\odot}$ for all lensing systems except for NFW and ENFW lens cases.
For the two lens models, the strong lensing cross sections ($\sigma_{1.0}$) with the above setting are too narrow and thus, we inject a GW150914-like event with $d_{L}=10~\rm{Gpc}$ and redshifted lens mass value $M^{z}_{l}=10^{13}\rm{M}_{\odot}$.

We use \textsc{IMRPhenomXPHM}~\citep{pratten2021computationally} to generate $h_{\rm{u}}$ and set possible source-lens systems, from double-image to quintuple-image systems where the number of the image is determined by the model-dependent parameters.
Considering the detectability of the LIGO–Virgo detectors~\citep{abbott2020prospects}, we assume that all lensed GW signals ($h_{l}(f)$) are detected at network signal-to-noise ratios (SNR) $\rho_{\rm net} > 8$ except for NFW and ENFW cases.\footnote{Type III images of NFW and ENFW are primarily demagnified, resulting in low SNRs for lensed GW signals.}
For each system, we choose the injected $\boldsymbol{\theta_{l}}$ shown in Table.~\ref{tab:inj_lenparams} to generate multiple lensed signals.
The SNR ranges of lensed GW signals are $\rho_{\rm net} = (11,16)$ for Systems 1,2 and 4-7, and $\rho_{\rm net} = (4,10)$ for Systems 3, 8 and 9\footnote{We set the SNR threshold as $\rho_{\rm net} =4$ for the demagnified images, which are expected to be detected through sub-threshold search~\citep{abbott2021search,abbott2023search,li2023targeted}.}.

\begin{table}[ht]
\begin{tabular*}{\linewidth}
{@{\extracolsep{\fill}} llll}
\toprule
\textbf{ID} &
\textbf{Model} & \textbf{Injected $\boldsymbol{\theta_{l}}$} & \textbf{No. of images} \\
\hline
1 & PM & $y = 0.22$ & Double \\
2& SIS & $y = 0.22$ & Double \\
3 &NFW & $(y,~\kappa_{s}) = (0.004,~0.167)$ & Triple \\ 
4 &SIE & $(q,~y) = (0.85,~0.22)$ & Double \\
5 &SIE & $(q,~y) = (0.63,~0.22)$ & Quadruple \\
6 &SPEMD & $(q,~y,~\gamma) = (0.85,~0.22,~1.5)$ & Double \\
7 &SPEMD & $(q,~y,~\gamma) = (0.58,~0.17,~0.75)$ & Quadruple \\
8 &ENFW & $(q,~y,~\kappa_{s}) = (0.6,~0.011,~0.167)$ & Triple \\
9 &ENFW & $(q,~y,~\kappa_{s}) = (0.82,~0.008,~0.167)$ & Quintuple \\
        \hline
        \hline
    \end{tabular*}
    \caption{Injected $\boldsymbol{\theta_{l}}$ values used to generate the nine GW lensing systems considered in this work.
    The parameters are randomly chosen from a pool of $\boldsymbol{\theta_{l}}$ that can generate the specified detectable multiple (double to quintuple) images.
    See Sec.~\ref{lensmodel} for each lens model's definition of lens parameters.
    \label{tab:inj_lenparams}}
\end{table}
Next, we infer the relative magnification factors $\mu_{\rm{rel}}$ and time delays $\Delta t$ among the lensed GW signals by performing joint parameter estimation on pairs of the first lensed signal ($h_{l,1}$) and the $j$-th lensed signal ($h_{l,j}$).
See Eq.~13 of~\cite{janquart2023return} for the joint likelihood function used in parameter estimation.
If the system has $n$ lensed GW signals, the relative magnification factors between $h_{l,1}$ and $h_{l,j}$ can be written as
\begin{equation}
    \mu_{\rm{rel},1\it{j}} = \frac{\mu_{j}}{\mu_{1}} = \left(\frac{d_{L,1}}{d_{L,j}}\right)^{2}
    \text{for} \quad j=2,...,n~,
\end{equation}
where $d_{L,j}$ is the apparent luminosity distance inferred from the $j$-th lensed signal.
Also, the actual time delays can be written as
\begin{equation}
    \Delta t_{1j} = |t_{c,j} -t_{c,1}| \quad \text{for} \quad j=2,...,n~,
\end{equation}
where $t_{c,j}$ is the apparent merger time inferred from the $j$-th lensed signal.

\begin{figure}[b]
    \centering
    \includegraphics[width=1.0\linewidth]{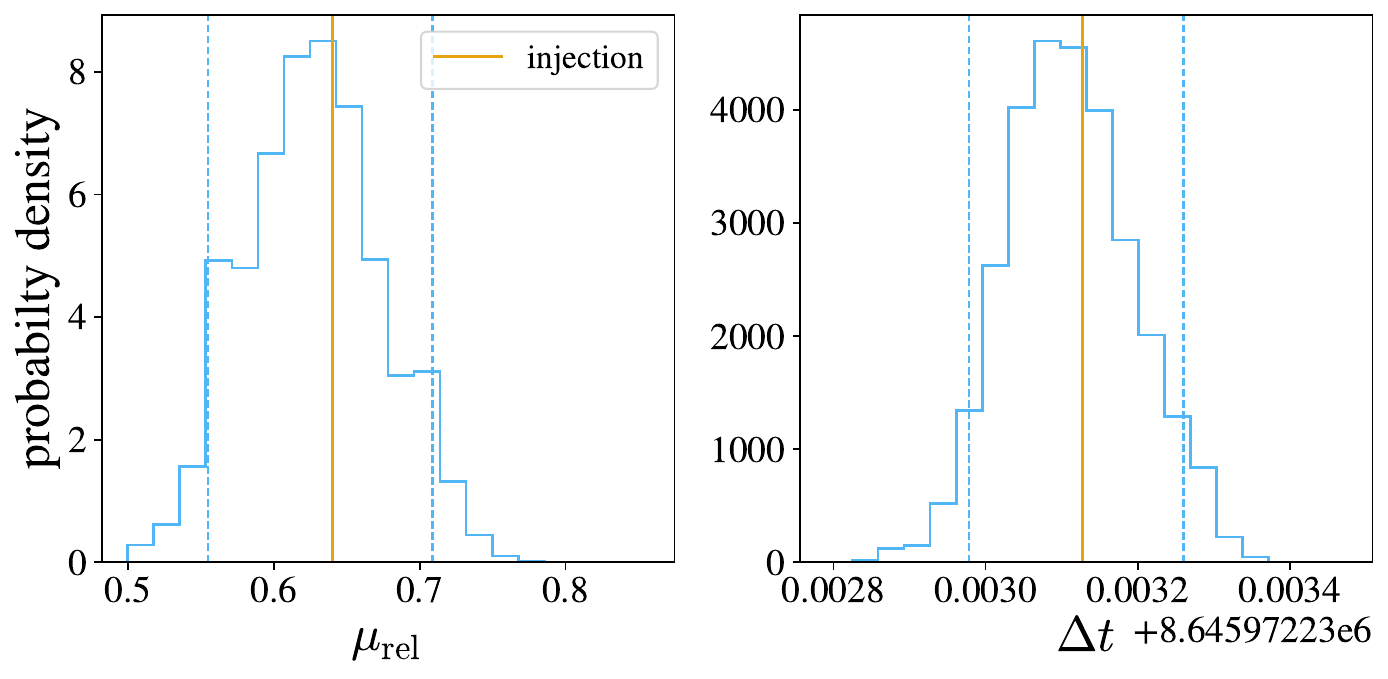}
    \caption{The posteriors of $\mu_{\rm rel}$ and $\Delta t$ from the joint parameter estimation for two lensed GW signals.
    An isolated point mass with $(y,M^{z}_{l})=(0.22, 10^{12}M_{\odot})$ is assumed.
    Gold lines indicate the true values.
    Vertical dashed lines mark the upper and lower bound of 90\% credible intervals.
    The lensing observables are well constrained, and the posteriors peak at the true values ($\mu_{\rm rel} \simeq 0.64$ and $\Delta t \simeq  8645972.23\rm{s}$).}
    \label{lensobserv}
\end{figure}
To conduct joint parameter estimation, we use the \textsc{golum} pipeline~\citep{janquart2021fast,janquart2023return} which in turn makes use of the Bayesian inference tools \textsc{bilby}~\citep{ashton2019bilby,romero2020bayesian} and \textsc{dynesty}~\citep{speagle2020dynesty} for nested sampling.
Simulated lensed GW signals are injected into Gaussian noise characterized by the representative LIGO-Virgo O4 power spectral density (PSD)~\citep{abbott2020prospects}.
We assume uniform prior distributions for the relative magnification factor $\mu_{\rm rel}$ and actual time delay $\Delta t$.
For the source parameters, we adopt the prior distributions for precessing binary black holes in ~\cite{romero2020bayesian}.
Priors of the main parameters used in this work are shown in Table.~\ref{tab:prior}.

\begin{table}[t]
\begin{tabular*}{\linewidth}
{@{\extracolsep{\fill}} llll}
\toprule
\textbf{Parameter} &
\textbf{Prior} & \textbf{Range} \\
\hline
$m_{1},m_{2}$ & \texttt{Uniform} & [5,~100]$M_{\odot}$  \\
$d_{L}$& \texttt{UniformComovingVolum}& [0.1,~5]Gpc OR  \\
& & [0.1,~10]Gpc \\
RA&\texttt{Uniform} & [0, 2$\pi$]  \\ 
DEC &\texttt{Isotropic} & -  \\
$\mu_{\rm rel}$ &\texttt{Uniform}& [0.1,10]  \\
$\Delta t$ &\texttt{Uniform} & - \\
        \hline
        \hline
    \end{tabular*}
    \caption{Prior assumptions for main source parameters, including component masses ($m_1,m_2$), luminosity distance ($D_{L}$), and sky location (RA, DEC), as well as lensing observables ($\mu_{\rm rel}$, $\Delta t$) employed in this work.
    A broader prior range for $D_L$ ([0.1, 10] Gpc) is adopted for NFW and ENFW lensing systems, reflecting injected $D_L$ values greater than 5~Gpc in these systems
    \label{tab:prior}}
\end{table}
Fig.~\ref{lensobserv} shows 
the posteriors of the $\mu_{\rm rel}$ and $\Delta t$ from the lensing system assuming the SIS model (see Table.~\ref{tab:inj_lenparams}).
The posteriors peak at the injected values with well-constrained credible intervals.
Although the peaks for the other lensing system cases cannot be precisely at the injected values due to the effects of noise, we confirm that the peaks of the posterior become nearer to the injected values when we use zero noise.
We will use these posteriors of $\mu_{\rm rel}$ and $\Delta t$ as constraints to infer the $\boldsymbol{\theta_{l}}$.

Except for a few simple models like PM and SIS lenses, the $\mu_{\rm rel}$ and $\Delta t$ do not have one-to-one relationships with $\boldsymbol{\theta_{l}}$.
Furthermore, the inferred $\mu_{\rm rel}$ takes the form of a probability distribution rather than a point estimate with a specific value, which may hinder us from accurately recovering the lens parameters.
In this sense, we can use the rejection sampling technique~\citep{mackay2003information} to infer the $\boldsymbol{\theta_{l}}$ since we can specifically determine the value of $\mu_{\rm rel}$ from a given $\boldsymbol{\theta_{l}}$ and have the probability distribution of $\mu_{\rm rel}$.

\begin{figure}[b]
    \centering
    \includegraphics[width=1.0\linewidth]{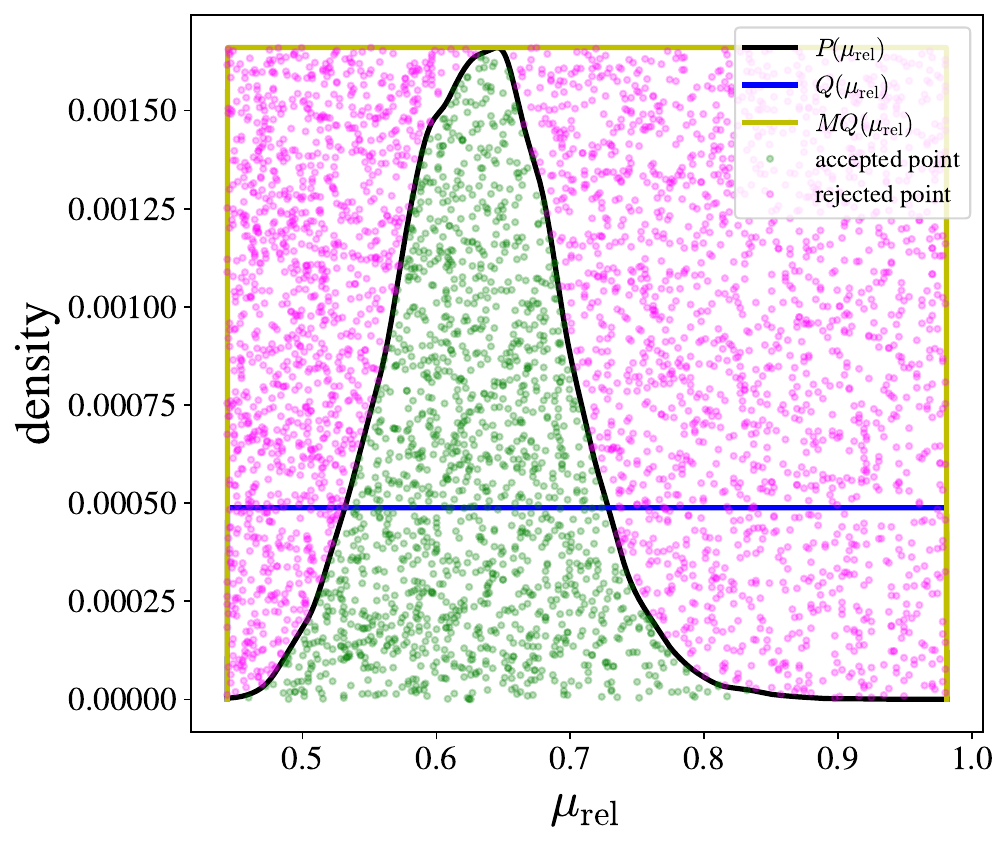}
    \caption{Illustration of rejection sampling.
    The solid blue line is the proposal distribution (uniform) to sample points, and the solid black line is the target distribution $P(\mu_{\rm rel})$ equal to the posterior of obtained $\mu_{\rm rel}$ in Figure~\ref{lensobserv}.
    The solid yellow line is a new proposal distribution made by multiplying a constant $M$ by the initial uniform distribution $Q(\mu_{\rm rel})$, which satisfies with $P(\mu_{\rm rel}) \leq MQ(\mu_{\rm rel})$.
    Based on the acceptance criterion, green and magenta points indicate the accepted and rejected $\boldsymbol{\theta_{l}}$, respectively.}
    \label{rej_sam}
\end{figure}

We use the lensing observable bank introduced in Sec~\ref{lensmodel}, which contains ordered triplets ($\Theta =\{\boldsymbol{\theta_{l}}, \mu_{\rm rel}, \Delta T\}$), in which the $\boldsymbol{\theta_{l}}$ create at least two lensed GW signals.
In this study, we use \textsc{lenstronomy}~\citep{birrer2018lenstronomy} to calculate $\mu_{\rm rel}$ and $\Delta T$ from a given $\boldsymbol{\theta_{l}}$, applying the rejection sampling method.
Thus, each sampled $\mu_{\rm rel}$ value is associated with a specific set of lensing parameter values, $\boldsymbol{\theta_{l}}$.
Fig.~\ref{rej_sam} shows an example case for the posterior distribution $P(\mu_{\rm rel})$, displaying accepted (green) and rejected (red) points following the application of the rejection sampling method.

Finally, the $\boldsymbol{\theta_{L}}$ points from the accepted triplets (green dots) may be used to represent the posteriors of these parameters.
Also, the associated $\Delta T$ points may be converted to the corresponding $M^{z}_{l}$ samples by using Eq.~\ref{mlzfromt}.

It is worth noting that the computation time of the rejection sampling technique is less than 1 CPU hour for a double-image system (first and second image pair).
For triple to quintuple-image systems, using rejection sampling would require a few CPU hours as it involves conducting rejection sampling for the first and $j^{\rm th}$ image pair only after obtaining the result for the first and $i^{\rm th}$ pair ($i<j$).
\section{Results}
\subsection{Axially symmetric lens cases}
\begin{figure}[t]
    \centering    \includegraphics[width=1.0\linewidth]{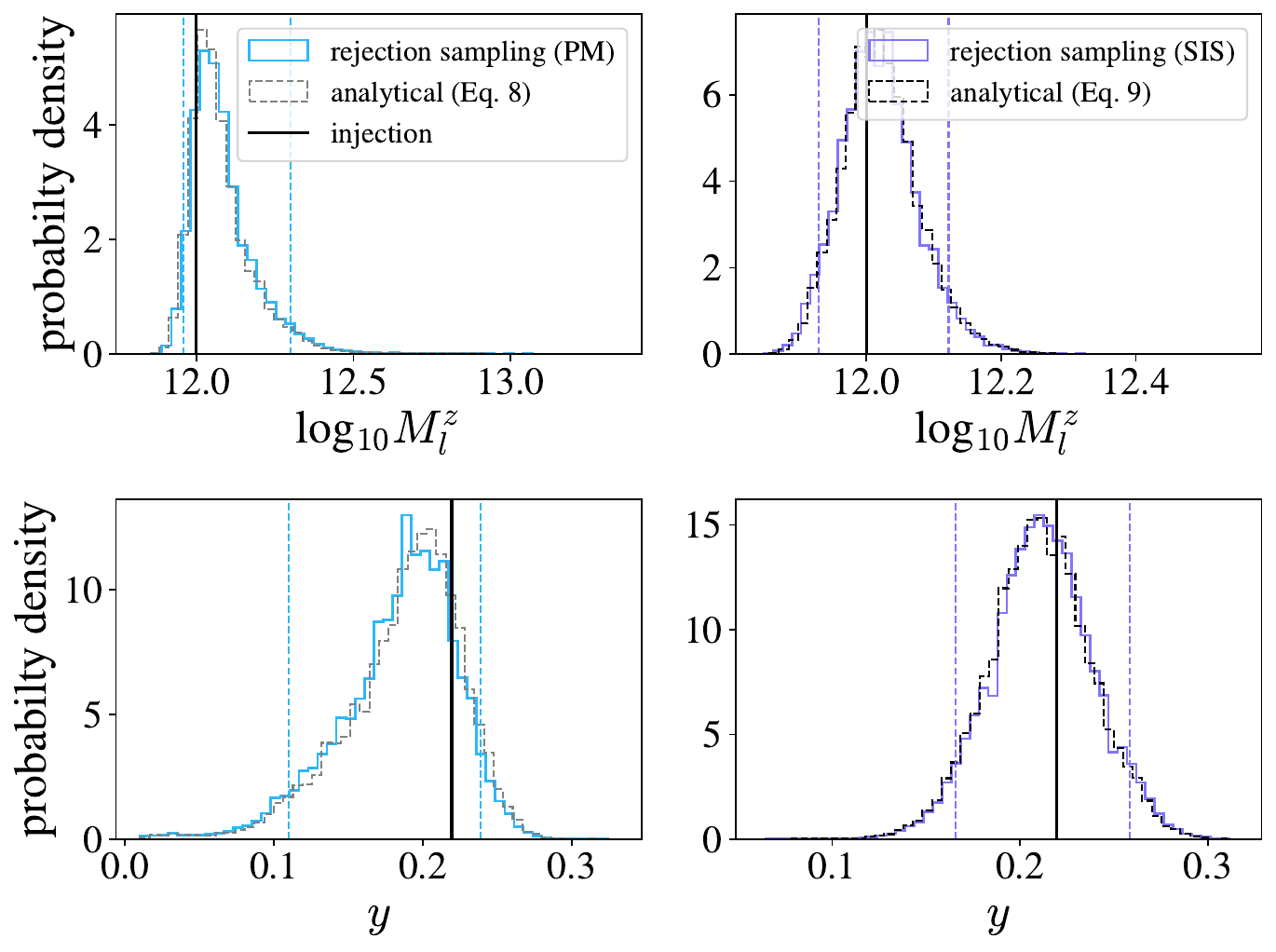}
    \caption{Posteriors of $\boldsymbol{\theta_{l}}$ for Systems 1 (PM double) and 2 (SIS double) in Table~\ref{tab:inj_lenparams} inferred from the relative magnification factor $\mu_{\rm rel}$ using rejection sampling (blue and violet).
    The solid black lines indicate that the injected values and the vertical dashed lines are 90\% symmetric credible intervals for each parameter.
    The $\boldsymbol{\theta_{l}}$ posteriors analytically obtained using Eq.~\ref{pm} and Eq.~\ref{sis} are shown with dashed grey and black lines, respectively.
    In both cases, the posteriors of $y$ and $M^{z}_{L}$ inferred using rejection sampling and analytically are almost identical, which demonstrates the validity of the rejection sampling technique.}
    \label{validity_check}
\end{figure}

To check the validity of the rejection sampling technique, we first compare the $\boldsymbol{\theta_{l}}$ posteriors inferred from rejection sampling to analytically obtained posteriors using Eqs.~\ref{pm} and~\ref{sis} for PM and SIS respectively.
For these two lens models, the relative magnification factors $\mu_{\rm rel}$ are univariate functions of the lens parameter $\boldsymbol{\theta_{l}}=y$.
Fig.~\ref{validity_check} shows that the recovered posteriors of $M^{z}_{l}$ and $y$ for System 1 (PM) and System 2 (SIS) using the above two methods peak at the injected values with great accuracy, and they are almost identical.
These results demonstrate that rejection sampling can be used to recover the lens parameters from the lensing observables.
Note, however, that this does not mean rejection sampling can successfully recover {\em all\/} $\boldsymbol{\theta_{l}}$ regardless of the degeneracies between them. 

Among the spherically symmetric lens models considered, the NFW model does not have an analytic expression for the lensing observables.
Thus, we use the rejection sampling technique for System 3 (NFW triple).
After identifying the estimated $\mu_{\rm rel}$ posteriors with the detected triple images as a target distribution, we draw $\mu_{\rm rel}$ samples from the lensing observable bank of the NFW to conduct rejection sampling.

\begin{figure}[t]
    \centering
\includegraphics[width=1.0\linewidth]{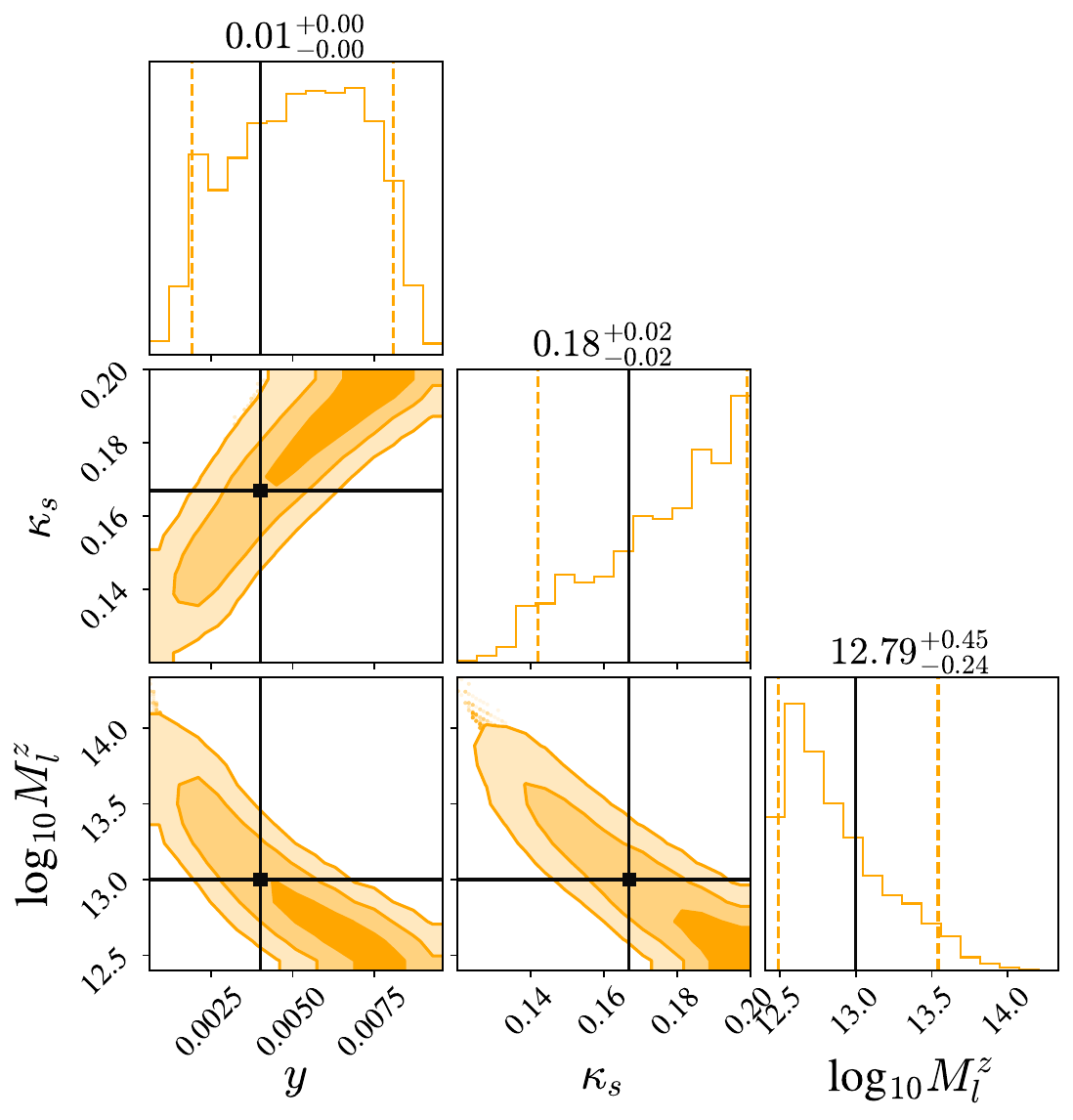}
    \caption{Corner plot showing the posteriors of the lens parameters $\boldsymbol{\theta_{l}}=(y,\kappa_{s})$ and redshifted lens mass $M^{z}_{l}$ for System 3.
    The solid black lines indicate the injected values, and vertical dashed lines mark the upper and lower bound of 90\% credible intervals.
    Each 1D posterior of $y$ and $\kappa_{s}$ is poorly constrained because the degeneracy between the two parameters is significant.
    Correspondingly, the $M^{z}_{l}$ does not peak at the true value.}
    \label{axially_result}
\end{figure}

Fig.~\ref{axially_result} shows the $\boldsymbol{\theta_{l}}$ and $M^{z}_{l}$ posteriors for the NFW case.
Even though injected values are contained within the 90\% credible intervals,  all posteriors are poorly constrained.
Even $\kappa_s$ is not converged well and does not change from the prior.
This is because many of the $(y,\kappa_{s})$ pairs in the prior pool, which can create multiple images, have similar $\mu_{\rm rel}$ values under galaxy-scale lens scenarios (i.e., only very small values of $y$ can generate GW strong lensing due to low surface density).
The obtained $\mu_{\rm rel}$ posterior is not well enough constrained to break this degeneracy between $y$ and $\kappa_s$.

Nevertheless, we can better recover the lens parameters of NFW when $y$ and $\kappa_s$ have large values (i.e., as will occur in galaxy cluster scale lens scenarios).

\subsection{Non-axially symmetric lens cases}
The ellipticity of a lens makes the lens model more complicated and degeneracies between $\boldsymbol{\theta_{l}}$ stronger.
Thus, the $\boldsymbol{\theta_{l}}$ and $M^{z}_{l}$ posteriors for non-axially symmetric lens models are more poorly constrained compared to those of spherically symmetric lens models.

\begin{figure*}[th]
    \centering
    \subfigure[System 4 (SIE double)]
    {
        \includegraphics[width=0.32\linewidth]{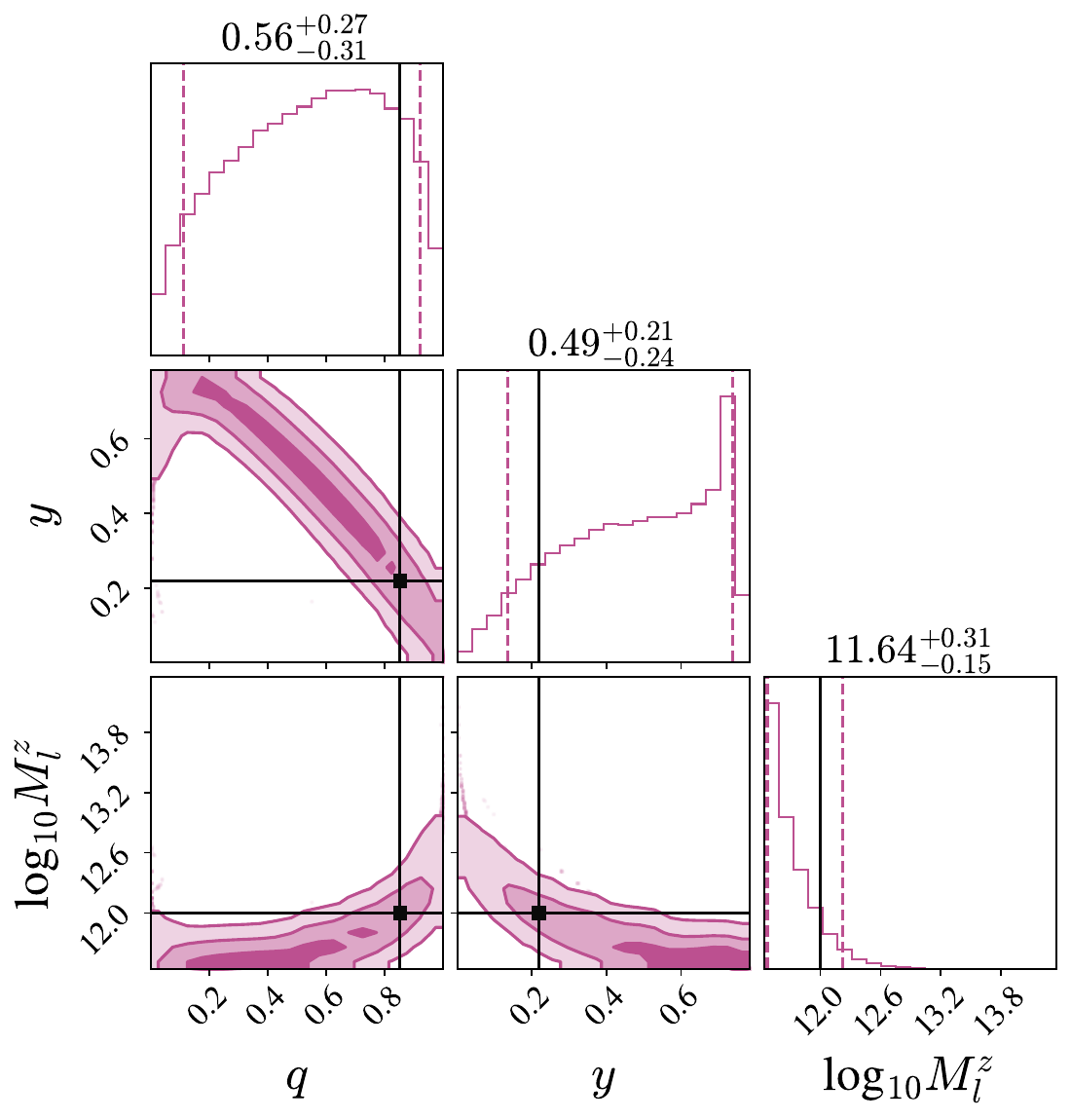}
        \label{sie_double}
    }
    \subfigure[System 6 (SPEMD double)]
    {
        \includegraphics[width=0.32\linewidth]{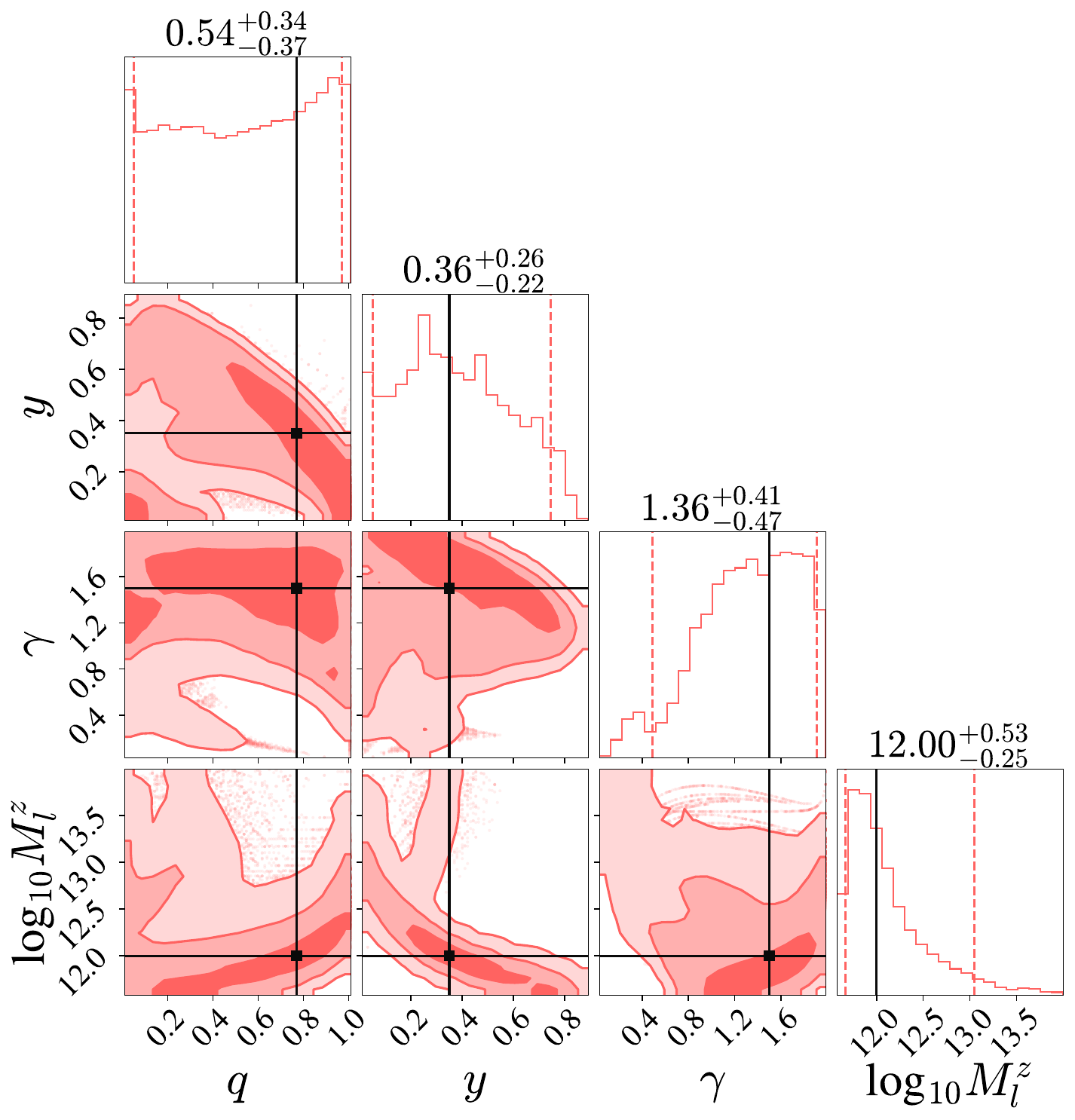}
        \label{spemd_double}
    }
    \subfigure[System 8 (ENFW triple)]
    {
        \includegraphics[width=0.32\linewidth]{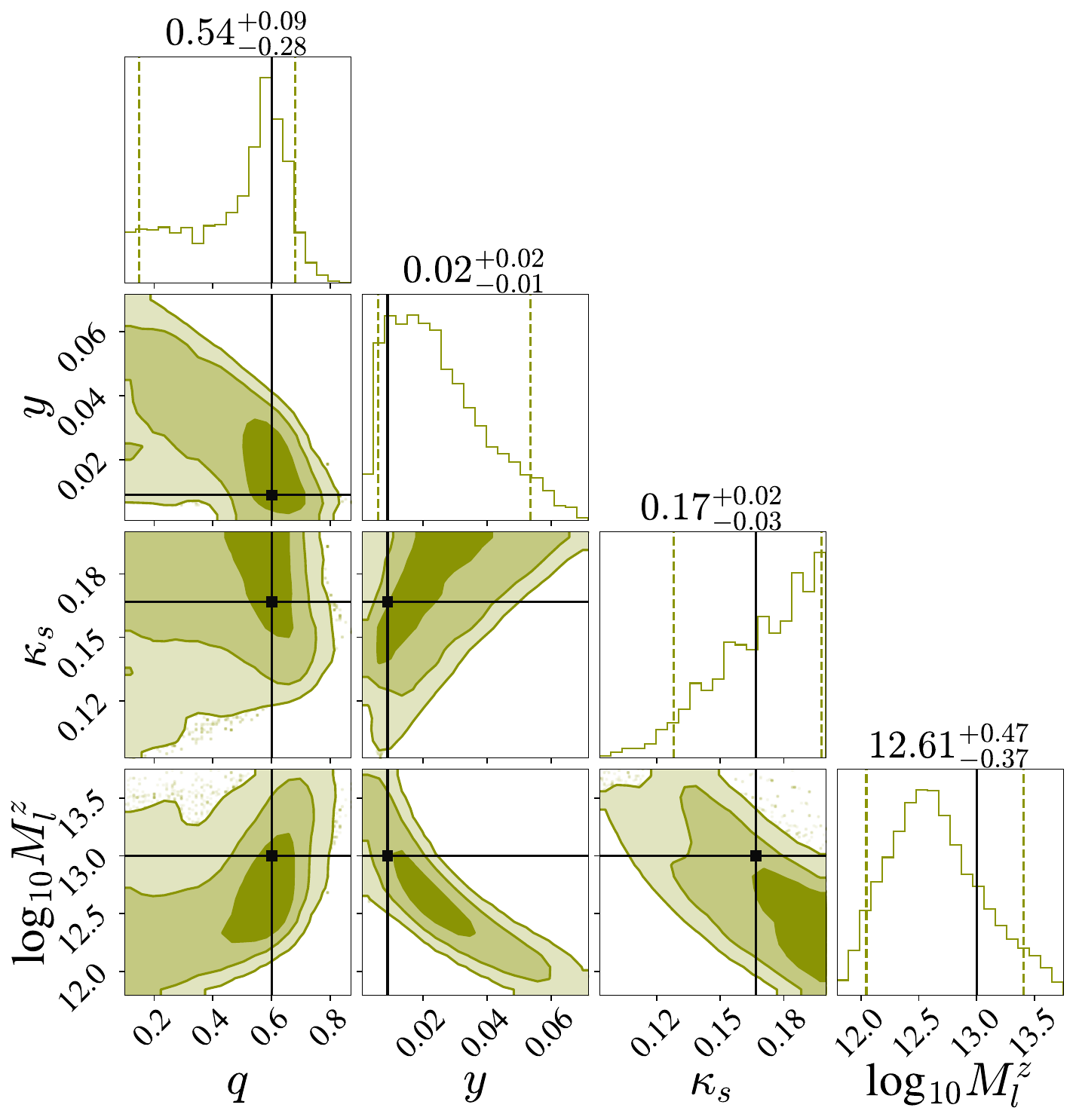}
        \label{enfw_triple}
    }
    \caption{Corner plots showing the posteriors of $\boldsymbol{\theta_{l}}$ and $M^{z}_{l}$ for Systems 4, 6, and 8.
    Solid black lines indicate the injected values and vertical dashed lines mark the upper and lower bound of 90\% credible intervals.
    Even though all injections are included within the 90\% credible intervals, the $\boldsymbol{\theta_{l}}$ posteriors do not converge well.
    Meanwhile, the $M^{z}_{l}$ posteriors are not highly biased from the injection because the mass strongly depends on the time delays between images, which can be constrained much better than relative magnification factors.
    Overall, there are limitations on accurately estimating dark halo properties with only double or triple-image systems.
    }
\label{non_axially_double_result}
\end{figure*}
Fig.~\ref{non_axially_double_result} shows the estimated posteriors of $\boldsymbol{\theta_{l}}$ and $M^{z}_{l}$ for Systems 4, 6, and 8 in Table \ref{tab:inj_lenparams} where double images (SIE and SPEMD) and triple images (ENFW) are detected.
Even though, for all three models, the true values of the parameters lie within the 2$\sigma$ credible regions of the joint 2-D posteriors of the lens parameters, the marginal 1-D posterior of each parameter does not peak at the injected value and is not well constrained.
This is because complex elliptical lenses have more lens parameters than spherically symmetric lenses, which results in stronger degeneracies between the parameters of $\boldsymbol{\theta_{l}}$.
Therefore, the properties of an elliptical dark galactic halo cannot be accurately estimated when the GW lensing system, having the halo, generates only double or triple images considering our lens model assumptions.

\begin{figure*}[th]
    \centering
    \subfigure[System 5 (SIE quadruple)]
    {
        \includegraphics[width=0.32\linewidth]{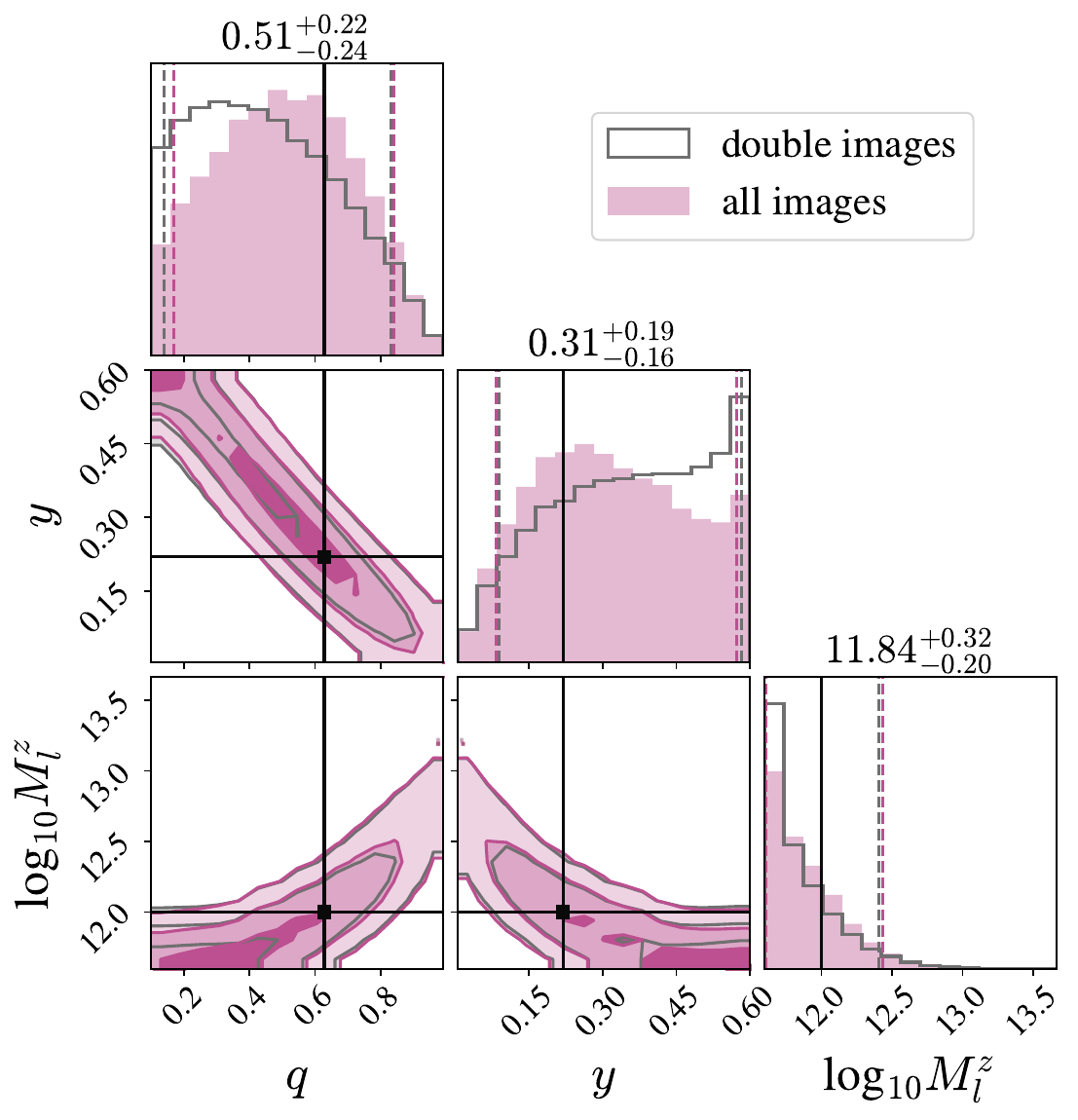}
        \label{sie_quadruple}
    }
    \subfigure[System 7 (SPEMD quadruple)]
    {
        \includegraphics[width=0.32\linewidth]{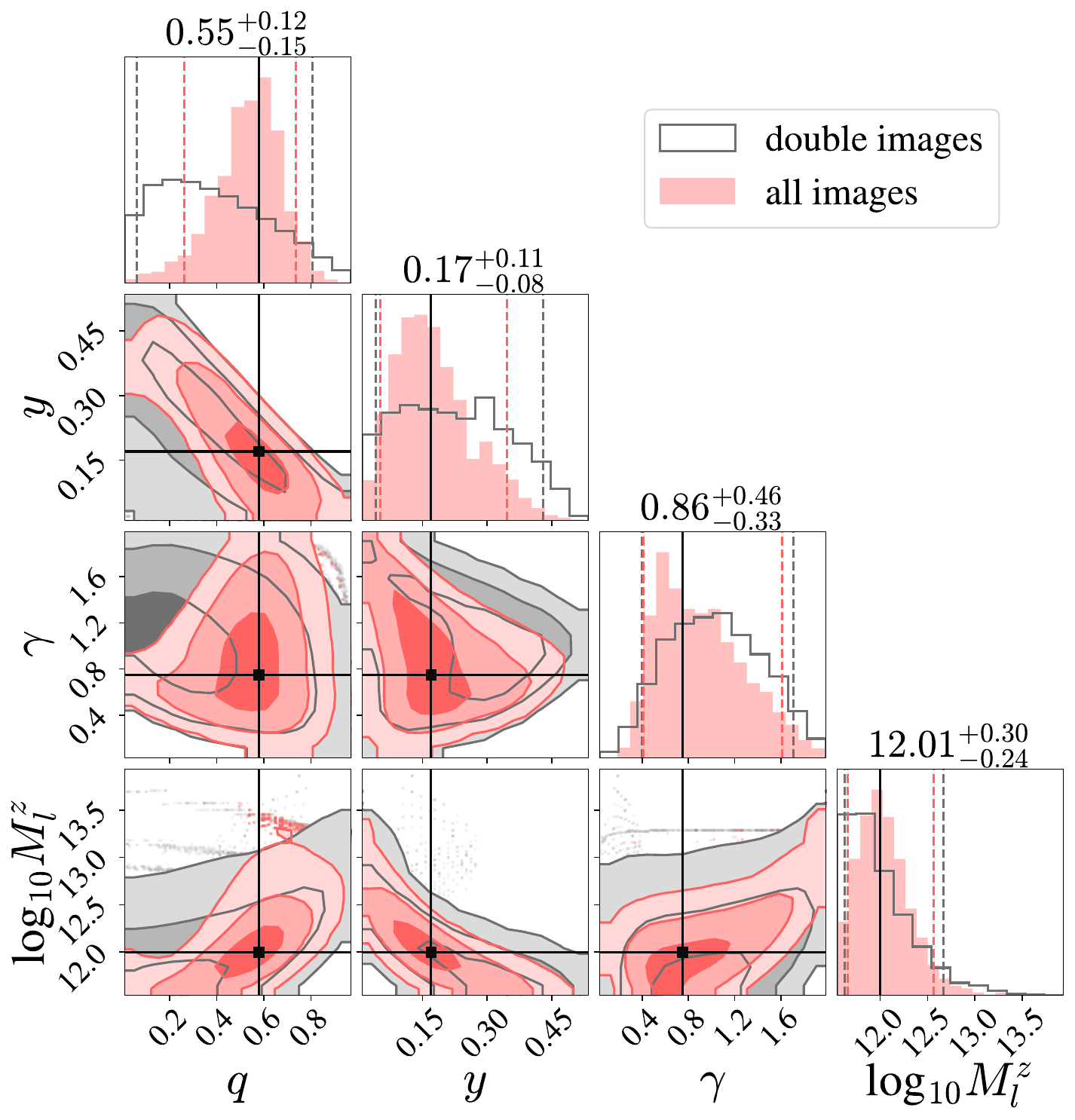}
        \label{spemd_quadruple}
    }
    \subfigure[System 9 (ENFW quintuple)]
    {
        \includegraphics[width=0.32\linewidth]{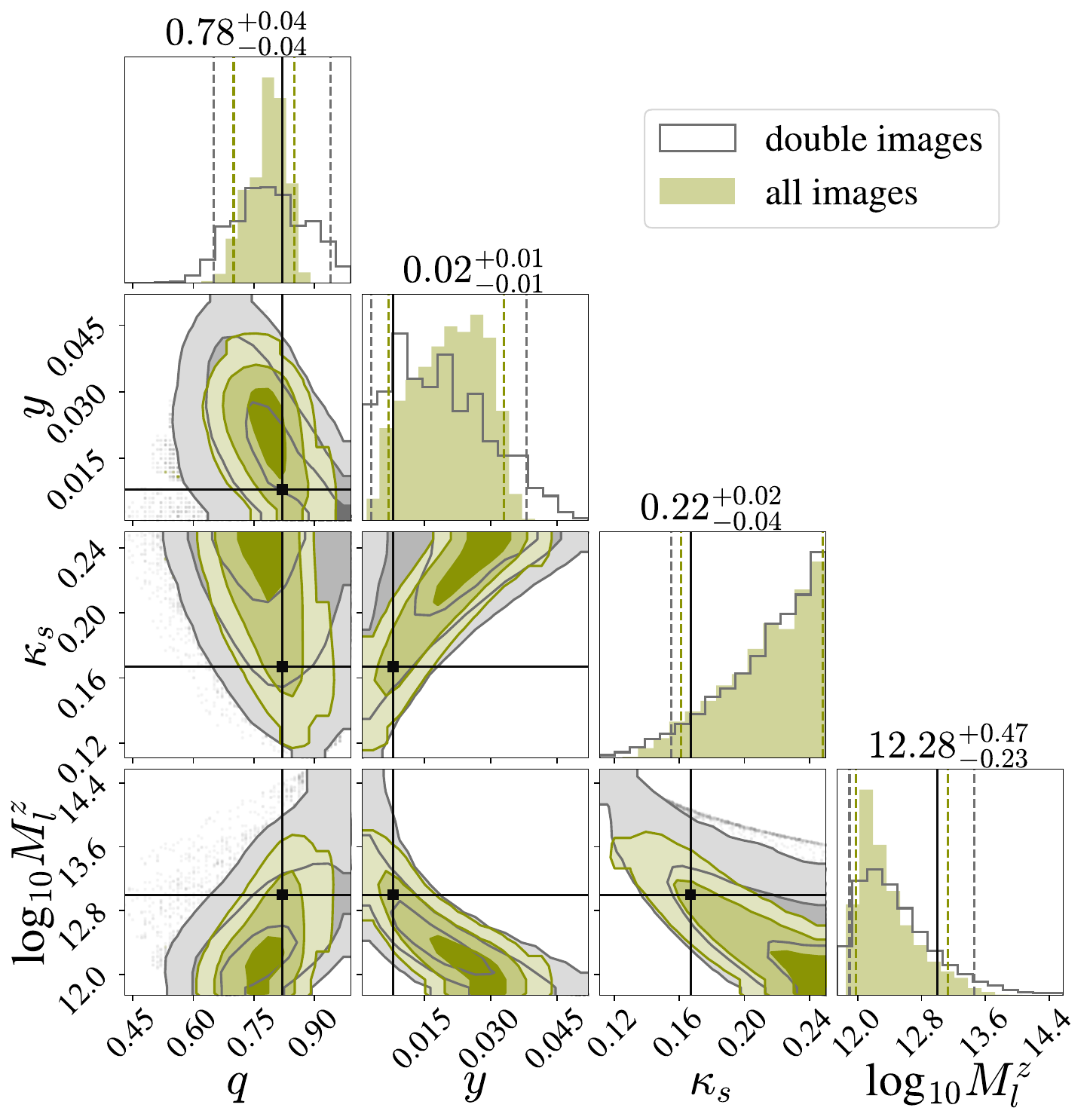}
        \label{enfw_quintuple}
    }
    \caption{Same configurations as in Fig.~\ref{non_axially_double_result}, but showing the posteriors of $\boldsymbol{\theta_{l}}$ and $M^{z}_{l}$ for Systems 5, 7, and 9.
    Histograms filled with colour represent the posteriors obtained from all sets of quadruple (or quintuple) GW images, whereas histograms depicted as solid grey lines illustrate the posteriors obtained exclusively from double images, which constitute a subset of the entire image collection.
    Since more constraints (i.e., more $\mu_{\rm rel}$  and $\Delta t$) are provided, the inferred posteriors for $\boldsymbol{\theta_{l}}$ are better constrained than those in the case where a double image is detected.
    Specifically, for all models, the widths of the posteriors for $q$ are narrowed down, which can be helpful to localize the lens galaxy in the inferred sky area.}
    \label{non_axially_quadruple_result}
\end{figure*}

When more than double images are created and detected, however, the degeneracies between $\boldsymbol{\theta_{l}}$ of elliptical lens models can be further reduced thanks to there being more constraints.
In Fig.~\ref{non_axially_quadruple_result}, we present the posteriors of $\boldsymbol{\theta_{l}}$ and $M^{z}_{l}$ for Systems 5, 7, and 9.
These systems produce quadruple or quintuple GW images, but it is plausible that we may only observe a subset of the complete GW image ensemble.
In light of this, we present the outcomes of two scenarios: 1) the detection of only pairs of GW images and (grey) 2) the detection of all GW images (coloured).

Table~\ref{tab:comparison} shows how much the recovery of lens parameters is improved for non-axially symmetric lenses when additional constraints are available.
We calculate a ratio $\mathcal{R}^{\theta_{l}}_{\rm{A/B}}$ between two 90\% credible intervals of $\theta_l$ posteriors obtained from case A and case B.
In reference to Fig.~\ref{non_axially_quadruple_result}, cases A and B denote scenarios where only two GW images are detected and all GW images are detected, respectively.
Except for the SIE case, the 90\% credible interval widths of $\boldsymbol{\theta_{l}}$ posteriors narrowed down by tens of percent.

Likewise, the posteriors obtained from more than double images show better results regarding the degree of convergence to the injected values. 
This is especially the case for SPEMD, where all $\boldsymbol{\theta_{l}}$ posteriors converge well to the injected values.
For the ENFW case, on the other hand, the degeneracy between $y$ and $\kappa_s$ is significant, similar to the NFW lens case.
Thus, for non-axially symmetric models, the posteriors of $\boldsymbol{\theta_{l}}$ and $M^{z}_{l}$ are generally not well constrained from GW information alone.

\subsection{Non-axially symmetric lens + indicative EM observation cases}
\begin{figure*}[t]
    \centering
    \includegraphics[width=0.33\linewidth]{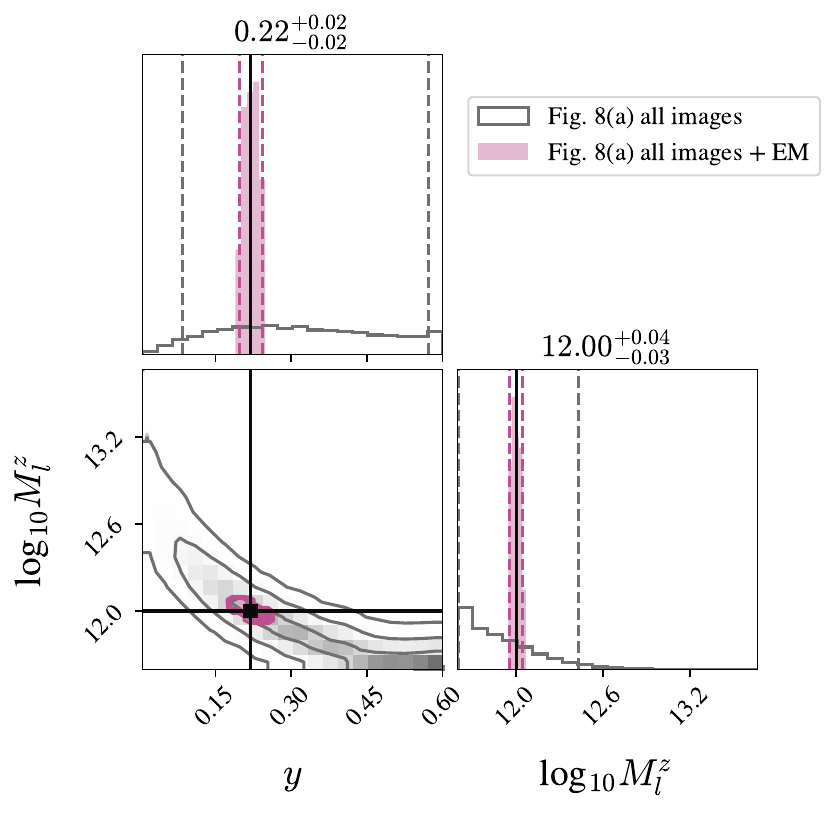}
    \includegraphics[width=0.33\linewidth]{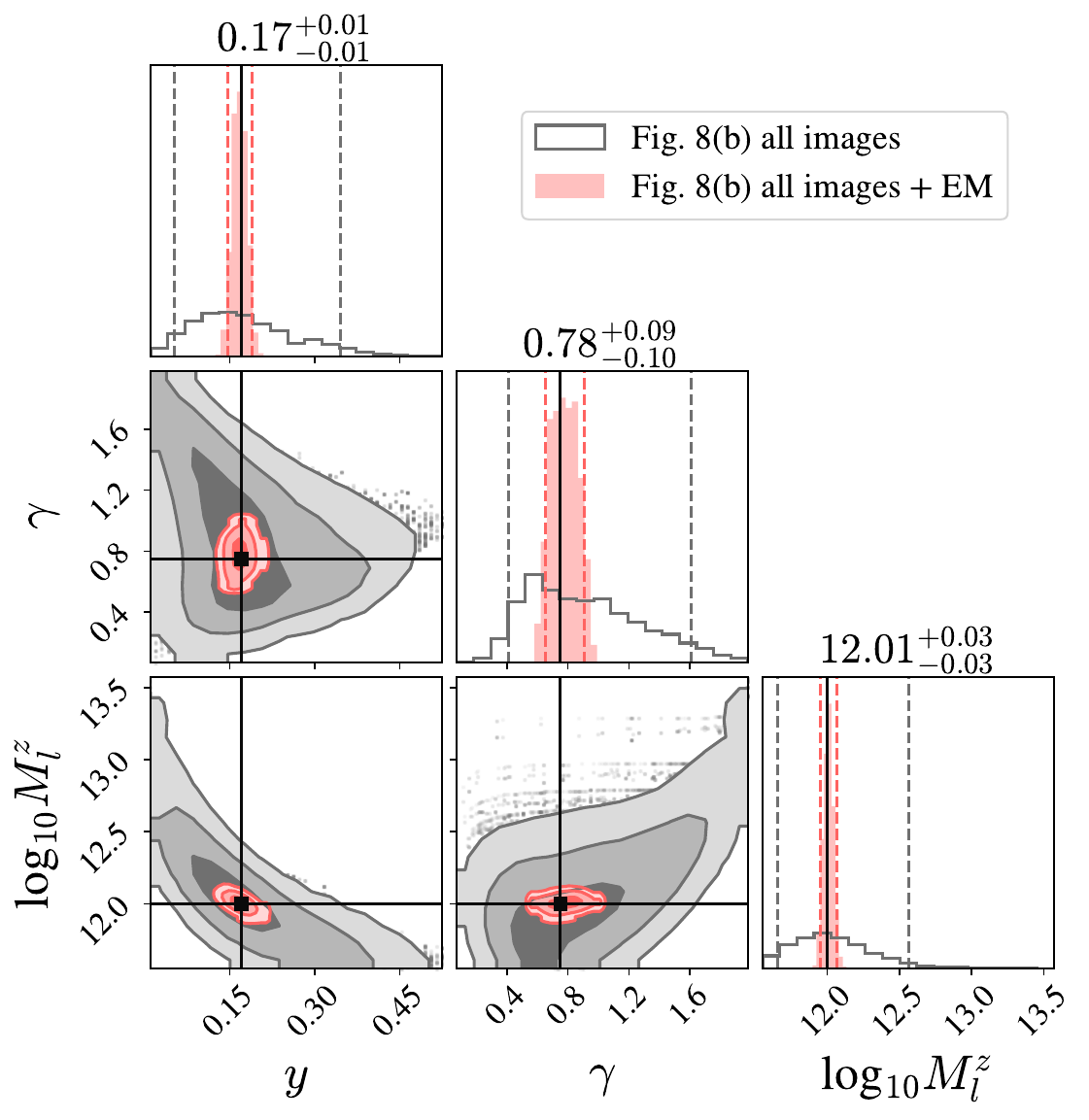}
    \includegraphics[width=0.33\linewidth]{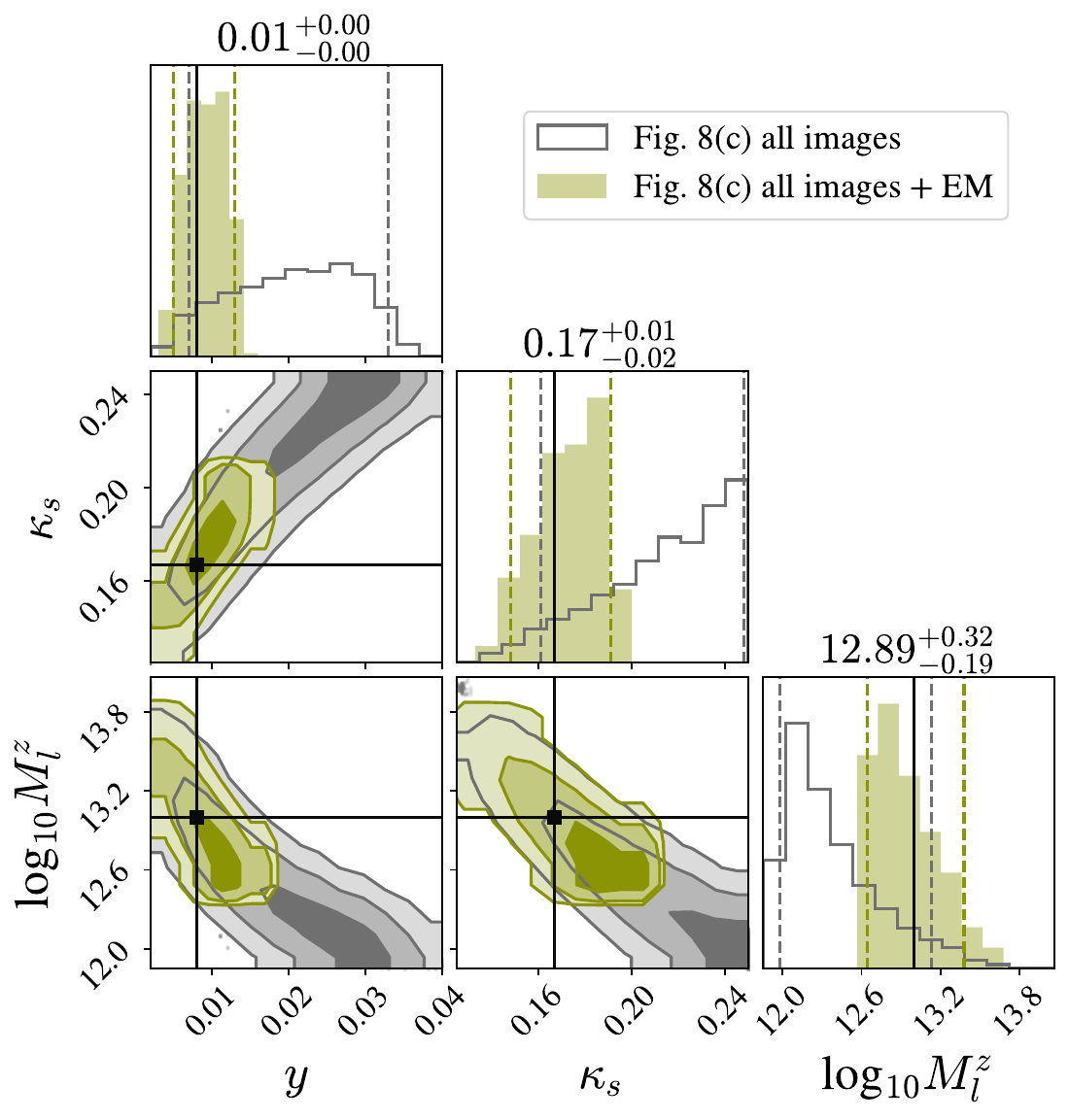}
    \caption{Same configurations as in Fig.~\ref{non_axially_quadruple_result}, but with the assumption that quadruple (quintuple) image systems are detected with the indicative electromagnetic observations and thereby, axis ratios and image positions are confined with $\pm$ 5\%  and $0.1^{"}$ error ranges.
    For easier comparison to the results without constraints from indicative EM observations, we plot the posteriors in Fig.~\ref{non_axially_quadruple_result} with solid grey lines.
    The inferred posteriors of $\boldsymbol{\theta_{l}}$ and $M^{z}_{l}$ for all non-axially symmetric lens models peak around the injected values (solid black lines) within 90\% credible intervals.
    Notably, the recovery of $M^{z}_{l}$ shows improvement, enabling more precise calculations of $M_{l}$ with the estimated $z_l$ of the host galaxy.}
    \label{non_axially_fixq_result}
\end{figure*}
In this sub-section, we consider how extra constraints from EM observations can reduce the degeneracies between parameters further, so that $M^{z}_{l}$ can be better constrained.
For instance, the axis ratio of the lens galaxy and EM image positions of the lensed host galaxy can be directly estimated if one explicitly identifies the lens galaxy.
However, it is challenging to search for the lens galaxy in the GW localisation area, which may contain numerous galaxies.

In this context, \cite{yu2020strong,hannuksela2020localizing,wempe2022lensing} propose that one can localize the host galaxy-lens galaxy system using LSST surveys~\citep{abell2009lsst} subject to several assumptions, such as the detection of both lensed GW signals and EM signals from the host galaxy\footnote{Generally, the host galaxy of a GW source can also be lensed by intervening dark galactic halos.}.
Based on the magnification factors and time delays obtained from the GW and EM observations, one can reduce the number of lens galaxy candidates within the sky area determined by the GW localisation.

In particular,~\cite{hannuksela2020localizing} argues that detecting quadruple GW images can effectively narrow down the sky area given the current LIGO-Virgo detectors.
Moreover, Fig.~\ref{non_axially_quadruple_result} shows that the axis ratio of a lens galaxy can be more precisely determined when four or more gravitational wave (GW) images are detected.
This GW information can be utilised to identify the lens galaxy with a well-constrained axis ratio.

Therefore, we assume here that the axis ratios of the lens galaxies and image positions can be confined to within an error range of $\pm$5\% and $0.1$ arcseconds, respectively, once each lens has been specified based on the observations of quadruple or more GW images.
Fig.~\ref{non_axially_fixq_result} shows the posteriors of $\boldsymbol{\theta_{l}}$ and $M^{z}_{l}$ of non-symmetric lens galaxies in each GW lensing system (See Table.~\ref{tab:inj_lenparams}) derived under the above assumption.
Thanks to the reduced degeneracies between $\boldsymbol{\theta_{l}}$, all their posteriors are much better constrained and peak at the injected values.
The corresponding $\mathcal{R}^{\theta_{l}}_{\rm{all} / \rm{all}+\rm{EM}}$ values are shown in Table~\ref{tab:comparison}.
These values are defined as the ratio of two 90\% credible intervals of $\theta_{l}$ posteriors obtained from the scenarios where all GW images are detected and where all GW images, along with indicative EM images, are detected.
Above all, the injected $M^{z}_{l}$ are successfully recovered.
One can calculate the lens mass $M_{l}$ with the inferred redshifted lens mass $M^{z}_{l}$ and estimated redshift $z_{l}$ from the corresponding EM observation.

\begin{table*}[t]
\begin{tabularx}{1.\linewidth}{@{}X*{6}{c} @{}}
    \toprule
        \multicolumn{6}{@{}l}{\textbf{$P_{\rm dbl}(\boldsymbol{\theta_{l}})$ vs $P_{\rm all}(\boldsymbol{\theta_{l}})$ in Fig.~\ref{non_axially_quadruple_result}}}\\
        \hline 
        Model & 
        $\mathcal{R}^{q}_{\rm{dbl} / \rm{all}}$ & 
        $\mathcal{R}^{y}_{\rm dbl / \rm{all}}$ &
        $\mathcal{R}^{{\rm log}_{10} M^{z}_{l}}_{\rm dbl / \rm{all}}$ &
        $\mathcal{R}^{\gamma}_{\rm{dbl} / \rm{all}}$ &
        $\mathcal{R}^{\kappa_{s}}_{\rm{dbl} / \rm{all}}$\\
        \hline
        SIE  & 1.03 &  1.01 & 0.97 &- &-\\
        SPEMD  &  1.55 &  1.33 & 1.22&1.11&-\\
        ENFW & 1.89 &  1.36 &1.30&- &1.06\\
        \hline
        \hline\\
        \hline
        \hline
        \multicolumn{5}{@{}l}{\textbf{$P_{\rm all}(\boldsymbol{\theta_{l}})$ vs $P_{\rm all + EM}(\boldsymbol{\theta_{l}})$ in Fig.~\ref{non_axially_fixq_result}}} \\
        \hline
        Model & 
        &
         $\mathcal{R}^{y}_{\rm{all} / \rm{all}+\rm{EM}}$ &
         $\mathcal{R}^{{\rm log}_{10}M^{z}_{l}}_{\rm all / \rm{all+EM}}$ &
        $\mathcal{R}^{\gamma}_{\rm{all} / \rm{all}+\rm{EM}}$&
        $\mathcal{R}^{\kappa_{s}}_{\rm{all} / \rm{all}+\rm{EM}}$\\
        \hline
        SIE &  &9.87 &  8.17 &-&-\\
        SPEMD &  & 6.94 &  7.90& 4.21&-\\
        ENFW & & 3.07 & 1.51&-&1.85\\
    \hline
    \hline
\end{tabularx}
\caption{Quantitative evaluations of how much additional constraints, such as more GW images and indicative EM observations, improve the recovery of lens parameters. 
In the top table, $\mathcal{R}^{\boldsymbol{\theta_{l}}}_{\rm dbl/all}$ denotes the ratios of the 90\% credible interval widths of $\boldsymbol{\theta_{l}}$ posteriors obtained solely from double GW images ($P_{\rm dbl}(\boldsymbol{\theta_{l}})$) to those from all GW images ($P_{\rm all}(\boldsymbol{\theta_{l}})$) in  Fig.~\ref{non_axially_quadruple_result}.
Similarly, in the bottom table, $\mathcal{R}^{\boldsymbol{\theta_{l}}}_{\rm all/all+EM}$ represents the ratios of the 90\% credible interval widths of $\boldsymbol{\theta_{l}}$ posteriors obtained from all GW images ($P_{\rm all}(\boldsymbol{\theta_{l}})$) to those from all GW images with additional indicative EM observations ($P_{\rm all+EM}(\boldsymbol{\theta_{l}})$) in Fig.~\ref{non_axially_fixq_result}.}
\label{tab:comparison}
\end{table*}
\section{Discussion}
Dark galactic halos are keys to understanding galaxy evolution, formation of large-scale structures, and constraining cosmological parameters.
Precisely measuring their masses and structures is essential. 
Gravitational lensing of GWs can significantly contribute to GW astronomy as increasing numbers of lensed GWs are observed in the near future by the next generations of detectors.

One can complement the 3D map of the mass distribution from lensed GW signals by constraining the halo mass function.
It is important to note that, for NFW and ENFW models, we can constrain both the lens mass $M^{z}_{l}$ and the dimensionless surface density $\kappa_{s}$, which are directly related to cosmological parameters.
The relationship between the two parameters can also be investigated to constrain the concentration ($c$) and critical density ($\rho_c$).

In this work, we have proposed how to estimate lens and source parameters from the lensed GW signals using various lens models.
We assume that all lensed GW signals are detected and the correct lens model is used to analyse the signals.
Our results demonstrate that we can retrieve lens parameters from the information contained in the lensed GW signals, even where these parameters do not have trivial relationships with each other.

Furthermore, we can make use of EM observations, which become viable when quadruple or more GW image systems are detected, to confine the axis ratios of the lenses and image positions, thereby enhancing the accuracy of their mass measurements.
Although only a BBH source is discussed in this work, we can also consider BNS and NSBH signals, where we could, in principle, observe both GW and EM lensing simultaneously, which facilitates localising the host galaxy~\citep{smith2023discovering,ma2023detection}.
In addition, while we used geometrical information of the EM observations, it is noteworthy that one can estimate the source redshift from lensed EM images using a spectroscopic method in order to calculate true luminosity distance to the source and individual magnification factors, providing additional constraints to the lens parameters. This is particularly possible as redshifts from lensed images are almost identical to the original source redshift under galaxy-scale strong lensing~\citep{wang2022instantaneous}.

However, we recognise that the situation could be more complicated in real GW detection scenarios.
Firstly, we may observe only a subset of all the lensed GW signals.
For example, only two lensed signals might be observed when an SIE galaxy generates four lensed signals from a given GW source. 
Assuming that the GW detectors keep operating and our lens model is accurate, there are two scenarios: i) The two non-detected signals have not yet arrived because of their longer time delays. ii) The two non-detected signals have relatively lower SNR than the two detected signals.
Therefore, one should perform rejection sampling assuming both double and quadruple-image triplets to investigate which scenario is more plausible and which number of images is preferred.

Next, we cannot instantly determine the lens density profile when we detect a set of lensed GW signals from a given GW source.
However, by reversely incorporating the rejection sampling technique into joint parameter estimation steps, one can conduct hypothesis testing to determine which model is more preferred to describe the lensed GW signals.
This process will also be helpful in the search for strongly lensed GW signals because \cite{ccalicskan2023lensing,janquart2023ordering} argue that lens model-dependent joint parameter estimations are necessary to overcome false alarms and improve the search.

Finally, we note that the substructure within a lens galaxy, such as star clusters and dark matter subhalos, could induce single or multiple microlensing signals to occur in a strongly-lensed GW event~\citep{diego2019observational,mishra2021gravitational,seo2022improving,yeung2023detectability}.
As a result, additional lens parameters would be required to describe the lensed GW signals, also necessitating the application of wave optics to fully understand the entire lensing effects.\footnote{When contemplating a space-based interferometer, which may enable us to observe significantly higher SNR events, the enhanced degeneracies among the lens parameters can in principle be resolved within the wave-optics regime.~\citep{tambalo2022gravitational,ccaliotacskan2023observability,ccalicskan2023probing}}
Since the rejection sampling technique is highly dependent on the magnification factors and time delays, ignoring these additional microlensing effects could result in incorrect estimation of the lens parameters.
Thus, the development of parameter estimation techniques that incorporate macrolens-microlens lensing systems is necessary. In future work, we will expand our study to address the above problems.
\section*{Acknowledgements}
The authors would like to thank Justin Janquart for carefully reading the manuscript and giving insightful comments.
E. S. is supported by grants from the Research Grants Council of the Hong Kong (Project  No. CUHK  24304317), the  Croucher Foundation of Hong Kong, the Research Committee of the Chinese University of Hong Kong, and the College of Science and Engineering of the University of Glasgow.
T. G. F. is supported by FWO (Research Foundation Flanders) Research Project nr. G086722N and FWO International Research Infrastructure nr. I002123N.
M. H. is supported by the Science and Technology Facilities Council (Grant Ref. ST/V005634/1).
The authors are grateful for computational resources provided by the LIGO Laboratory and supported by the National Science Foundation Grants PHY-0757058 and PHY-0823459
\newpage
\appendix
\section{Lensing formalism for axially-symmetric lens models}\label{appendix_simple}
\subsection{point mass (PM)}
Following the strong lensing configuration presented in Section~\ref{sec:2}, the image and source positions for a PM lens are normalised by the lens Einstein radius, given in Eq.~\ref{einstein_radius}, and the dimensionless lens equations for the PM model are given by
\begin{align}
\label{leq_pm}
    \vec{y} &= \vec{x} - \nabla\psi(\vec{x}) \nonumber\\
    &= \vec{x} - \nabla(\ln{\vec{x}}) \nonumber\\
    &= \vec{x} - \frac{1}{\vec{x}}
\end{align}
The number of solutions of Eq.~\ref{leq_pm} is always two, which means that two GW images are always created by a PM lens independent of the dimensionless source position $y$ (see Fig~\ref{simple_caustic}).
The magnification factors and Fermat potentials of the two GW images can be analytically calculated by using Eq.~\ref{eq:mu_fp}.
The caustic, where the magnification diverges, of a GW lensing system with a PM lens is a point $\vec{y}=(0,0)$.
Also, the lensing observables of the PM model, which can be obtained using Eqs.~\ref{eq:murel} and~\ref{eq:dt}, are expressed analytically as follows, 
\begin{eqnarray}
\label{pm}
\mu_{\mathrm{rel}}=\frac{2\left(y^2+2\right)}{y \sqrt{y^2+4}+y^2+2} -1\nonumber , \\
\Delta T=\frac{y \sqrt{y^2+4}}{2}+\ln \left(\frac{\sqrt{y^2+4}+y}{\sqrt{y^2+4}-y}\right).
\end{eqnarray}
\subsection{Singular isothermal sphere (SIS)}
The density profile of the SIS model is given by 
\begin{equation}
    \rho(r) = \frac{{\sigma_{v}}^{2}}{2\pi G r^{2}},
\end{equation}
where $\sigma_{v}$ is the velocity dispersion of the particles.
Typically, the Einstein radius of the SIS model is a function of $\sigma_{v}$,
\begin{equation}
\label{er_sis}
    r_{E,\rm{SIS}} = 4 \pi \left(\frac{{\sigma_{v}}}{c}\right)^{2}\frac{D_{L}D_{LS}}{D_{S}}~,
\end{equation}
which can be calculated using stellar dynamics, while the Einstein radius in Eq.~\ref{einstein_radius} was specified by a given Einstein mass.
These two definitions of the Einstein radius are basically the same in a gravitational lensing system.
To simulate the GW lensing system with the SIS model, we adopt Eq.~\ref{einstein_radius} as the normalisation constant, considering that lensing observables, namely, the time delays between multiple GW images ($\Delta t$), depend on Einstein mass.

Like the PM model, the dimensionless lens equations for the SIS model have an analytical form,
\begin{align}
\label{leq_sis}
    \vec{y} &= \vec{x} - \nabla\psi(\vec{x}) \nonumber\\
    &= \vec{x} - \nabla(\vec{x}) \nonumber\\
    &= \vec{x} - \frac{\vec{x}}{|x|}
\end{align}
Eq.~\ref{leq_sis} has two solutions if $|y|<1$, while it has a unique solution if $|y| \geq 1$ (see Fig.~\ref{simple_caustic}).
In the case of $|y|<1$, the relative magnification factor and the dimensionless time delay between two images are written as
\begin{eqnarray}
\label{sis}
\mu_{\rm rel}=\frac{1-y}{1+y} \nonumber \\
\Delta T = 2y.
\end{eqnarray}

\subsection{Navarro-Frenk-White profile (NFW)}
The density profile of the NFW model is described by
\begin{equation}
\rho(r)=\frac{\rho_{s}}{\frac{r}{r_{s}}\left(1+\frac{r}{r_{s}} \right)^{2}}~,
\end{equation}
where $r_{s}$ is the scale radius and $\rho_{s}$ is the characteristic density of the dark halo.
Since the total mass of the NFW halo is divergent, we adopt the virial radius of the halo $r_{\rm vir}$ as the extremity of the halo.
With the $r_{\rm vir}$ and the concentration defined as $c_{\rm vir}\equiv r_{\rm vir}/r_{s}$, the characteristic density is expressed by
\begin{equation}
\rho_{s} = \frac{M}{4 \pi r_{\text {vir}}^3 } \frac{c_{\text {vir }}^3}{\left[\ln \left(1+c_{\text {vir }}\right)-c_{\text {vir }} /\left(1+c_{\text {vir }}\right)\right]}~,
\end{equation}
where $M$ is the total mass of the halo enclosed by $r_{\rm vir}$.

The dimensionless surface density $\kappa_{s}$ in Eq.~\ref{kappa} can be calculated with given $r_s$ and $\rho_s$, which describes the lensing effects due to the NFW profile.
By adopting the $r_{s}$ as the normalisation constant, the deflection angle $\alpha=\nabla \psi$ of the NFW model is given by
\begin{equation}
\label{def_ang_nfw}
\alpha_{\rm NFW}(x)=\frac{4\kappa_{s}}{x} \times \left\{\begin{array}{rl}
\ln \frac{x}{2}+\frac{2}{\sqrt{x^2-1}} \arctan \sqrt{\frac{x-1}{x+1}} & (x>1) \\
\ln \frac{x}{2}+\frac{2}{\sqrt{1-x^2}} \operatorname{arctanh} \sqrt{\frac{1-x}{1+x}} & (x<1) \\
\ln \frac{1}{2}+1 & (x=1)
\end{array} ,\right.
\end{equation}
where the image position $x=|\vec{\xi}|/r_{s}$.

\begin{figure*}[t]
    \centering
    \includegraphics[width=0.33\linewidth]{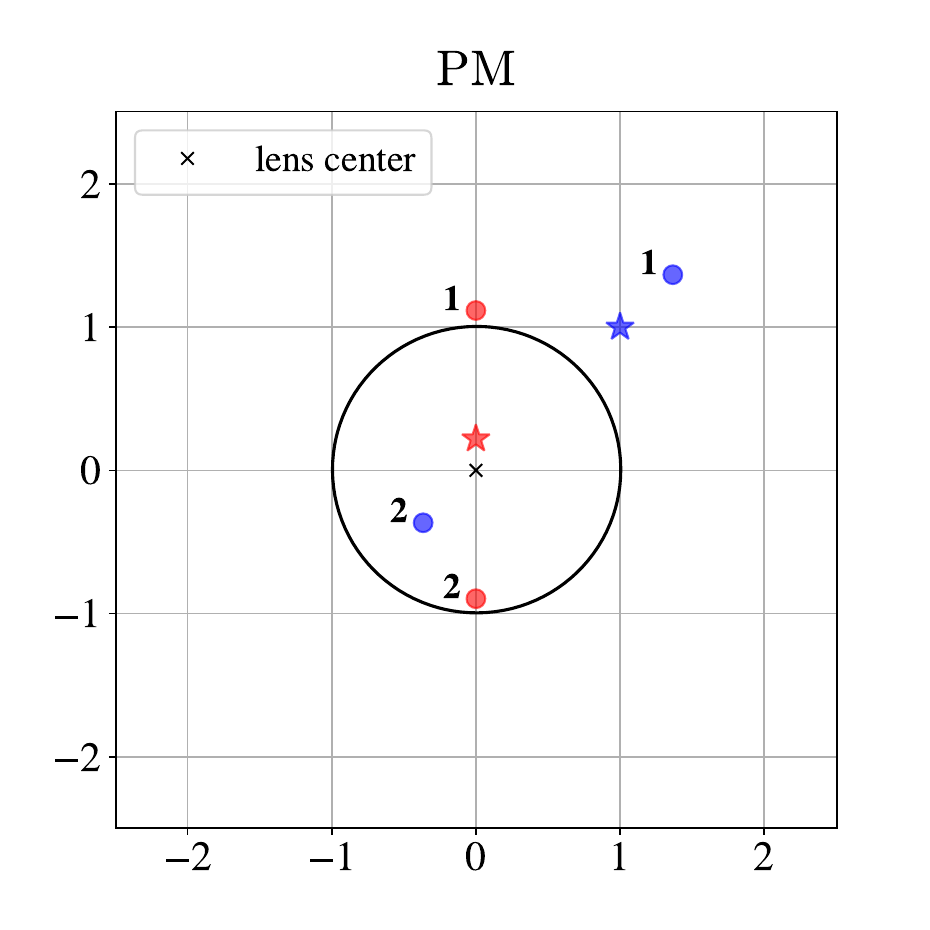}
    \includegraphics[width=0.33\linewidth]{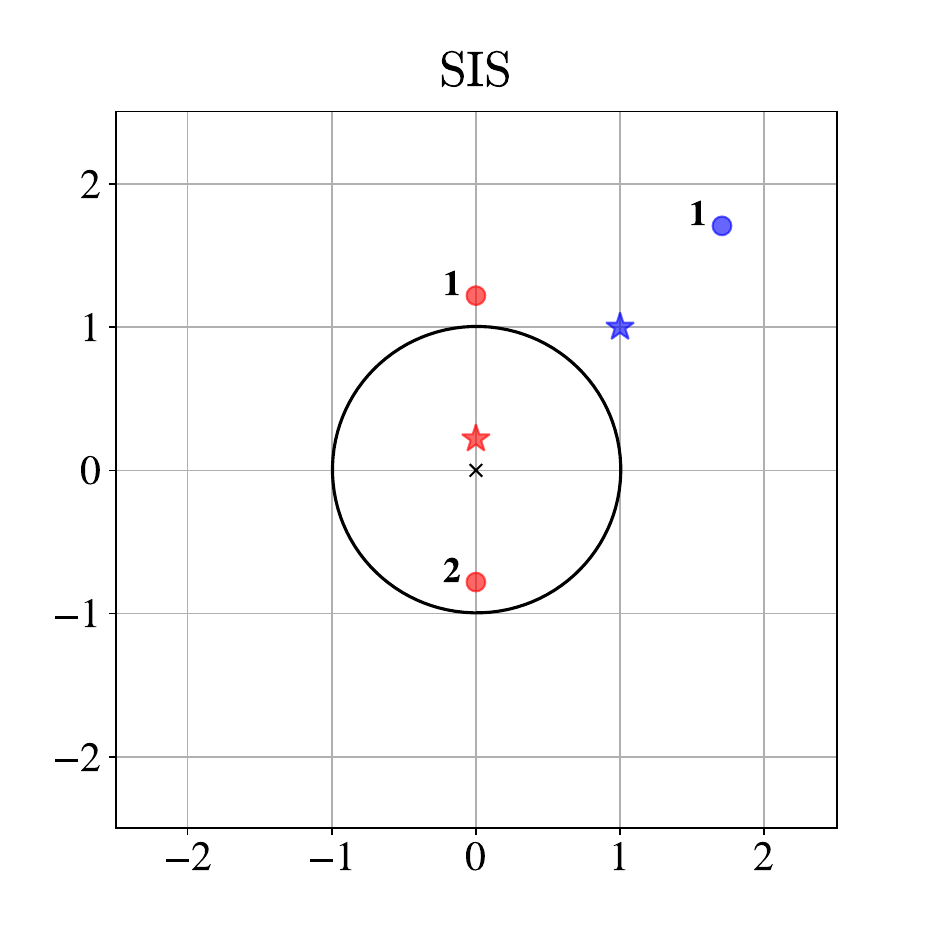}
    \includegraphics[width=0.33\linewidth]{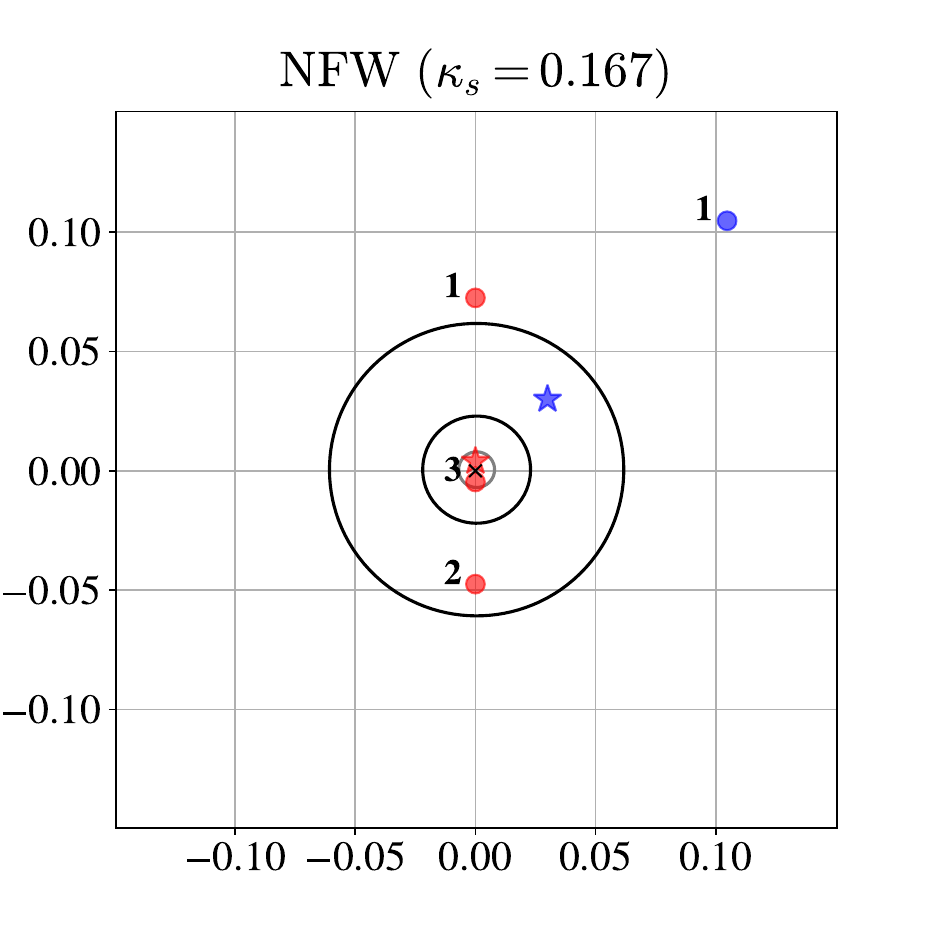}
    \caption{Critical lines (solid black lines) and caustics (solid grey lines) of PM, SIS, and NFW lenses specified in Table~\ref{tab:inj_lenparams} in the units of normalised image and source positions.
    Stars and dots indicate the sources and images respectively and in each case the set of images and the corresponding source are distinguished by colour.
    The numbers next to the images indicate the sequence in which the images arrive.
    The PM lens always creates two images regardless of the source position, while the SIS lens produces two images only if $y<1$.
    The NFW lens generates three images when a source is located within a caustic. 
    Note that the PM and SIS models have one critical line each because they do not have radial caustics.}
    \label{simple_caustic}
\end{figure*}

Unlike the above simple lenses, the NFW model does not have an analytic form for the lensing observables. 
Therefore, one needs to numerically derive the lensing observables using Eqs.~\ref{eq:murel} and~\ref{eq:dt}, i.e. using the deflection angle expression of~\ref{def_ang_nfw} to solve the lens equation. 
Fig.~\ref{simple_caustic} shows the critical liens and caustics of an NFW lens.
Similar to the SIS model, the lens equations for the NFW model have three solutions when $|y|<y_{cr}$, while a single solution is obtained for $|y|>y_{cr}$, where $y=y_{cr}$ indicates the radial caustic of the lens.

\section{Lensing formalism for complex lens model}\label{appendix_complex}
\subsection{Singular isothermal ellipsoid (SIE)}
Introducing ellipticity to the SIS model, the density profile of the SIE model is given by
\begin{equation}
\label{sie_density}
\rho(\xi_1,\xi_2)=\frac{{\sigma_{v}}^{2}}{2 \pi G} \frac{1}{\sqrt{\xi_1^2+q^2 \xi_2^2}}~,
\end{equation}
where $(\xi_1, \xi_2)$ is a position on the lens plane and $q$ is the axis-ratio of the lens.
By adopting the same Einstein radius used for the SIS model, the dimensionless deflection angle can be derived from Eq.~\ref{sie_density} (See.~\cite{kormann1994isothermal} for the detailed derivation),
\begin{equation}
\begin{split}
\label{def_ang_sie}
\alpha_{\rm SIE}(x_1,x_2)= &\sqrt{\frac{q}{1-q^2}}\left[\operatorname{arsinh}\left(\frac{\sqrt{1-q^2}}{q} \frac{x_1}{\sqrt{x^{2}_{1}+x^{2}_{2}}}\right)\right. \\
& + \left. \arcsin \left(\sqrt{1-q^2} \frac{x_2}{\sqrt{x^{2}_{1}+x^{2}_{2}}}\right) \right]~,
\end{split}
\end{equation}
where $x_{1}$ and $x_{2}$ are the $x$- and $y$- components of the image position.

In the same manner as for the NFW model case, one needs to solve the lens equation using the deflection angle in Eq.~\ref{def_ang_sie} to numerically obtain lensing observables for the SIE model.
Axial symmetry is broken for the SIE model, resulting in the transformation of the point (tangential) caustic at $\vec{y}=(0,0)$ of the SIS lens into a curved region with finite area.
Fig.~\ref{complex_caustic} illustrates the critical line, cut, and caustic of a SIE.
When a source is located within a cut, two images form; otherwise, a single image forms.
The cut encloses the caustic in this case, resulting in the creation of two additional images when a source is located within the caustic.
\subsection{Singular power-law ellipsoidal mass distribution (SPEMD)}
The SIE model can be further generalised by introducing a power-law form for its mass profile.
As mentioned in Sec.~\ref{ellip_lens}, we adopt the SPEMD model and its dimensionless convergence profile is written as
\begin{equation}
\label{spemd_profile}
    \kappa(x_1,x_2) = \frac{2-\gamma}{2}\left(\frac{q}{qx_1^2 + x_2^2}\right)^{\frac{\gamma}{2}},
\end{equation}
where $\gamma$ is the slope of the halo density.
Note that the Einstein radius of the SIS $r_{E,\rm{SIS}}$ is adopted as the normalisation constant.

Even though lensing effects due to the SPEMD must be calculated using numerical methods,~\cite{tessore2015elliptical} derived a pseudo-analytic form of the deflection angle of SPEMD with a complex formalism. Namely, $\alpha=\alpha_{1}+i\alpha_{2}$, where the real and imaginary parts indicate the first and second components of the deflection angle, respectively~\citep{bourassa1973spheroidal,bourassa1975theory}.
Assuming that $\alpha=\alpha_{x_{1}}+i\alpha_{x_{2}}$, one can derive the deflection angle at the image position $(x_1,x_2)$ from Eq.~\ref{spemd_profile} -- see Sec. 2 in \cite{tessore2015elliptical}, which is given by
\begin{equation}
\label{spemddeflecangle}
\begin{split}
\alpha_{\rm SPEMD}(x_1,x_2)= {} & \frac{2 \sqrt{q}}{1+q}\left(\frac{\sqrt{q}}{\sqrt{q^{2}x^2_{1}+x^{2}_{2}}}\right)^{\gamma-1} \\
& \times \mathrm{e}^{\mathrm{i} \varphi}{ }_{2} F_{1}\left(1, \frac{\gamma}{2} ; 2-\frac{\gamma}{2} ;-\frac{1-q}{1+q} \mathrm{e}^{\mathrm{i} 2 \varphi}\right)~,
\end{split}
\end{equation}
where $\varphi=\arctan(x_{2}/qx_{1})$ is the elliptical angle, and ${ }_{2} F_{1}$ is the Gaussian hypergeometric function.
Eq.~\ref{spemddeflecangle} can be used to solve the lens equations and to calculate the lensing observables.

When $\gamma \geq 1$, the SPEMD lens has a cut and a tangential caustic similar to the SIE lens.
Thus, double and quadruple images can be generated.
On the other hand, the cut disappears, and a radial caustic forms when $\gamma < 1$ because the singularity at the lens centre vanishes, and the lens produces triple or quintuple images.
Fig.~\ref{complex_caustic} shows how many images are formed depending on the source position for the SPEMD cases.

\subsection{Elliptical Navarro-Frenk-White profile (ENFW)}
Similar to the SIE case, NFW can be modified to ENFW by incorporating ellipticity.
We follow \citep{golse2002pseudo} introducing an elliptical coordinate system to compute the deflection angles of elliptical lens models.
We use  $x_{\epsilon}=\sqrt{x_{1\epsilon}^{2}+x_{2\epsilon}^{2}}=\sqrt{a_{1\epsilon}x_{1}^{2}+a_{2\epsilon}x_{2}^{2}}$ to express the deflection angle of ENFW as follows,
\begin{equation}
\begin{split}
\alpha_{\rm ENFW}(x_1, x_2)= {} &
\alpha_{\rm NFW}\left(x_\epsilon\right) \sqrt{a_{1 \epsilon}} \cos \phi_\epsilon \\
& + \alpha_{\rm NFW}\left(x_\epsilon\right) \sqrt{a_{2 \epsilon}} \sin \phi_\epsilon~,
\end{split}
\end{equation}
where $\phi_\epsilon = \arctan{(x_{2\epsilon}/x_{1\epsilon})}$ is the elliptical angle, similar to $\varphi$ in Eq.~\ref{spemddeflecangle}.
For consistency with the other elliptical lens models above, we define $a_{1\epsilon}= q$ and $a_{2\epsilon}= 1/q$.

\begin{figure*}[t]
    \centering
    \includegraphics[width=0.33\linewidth]{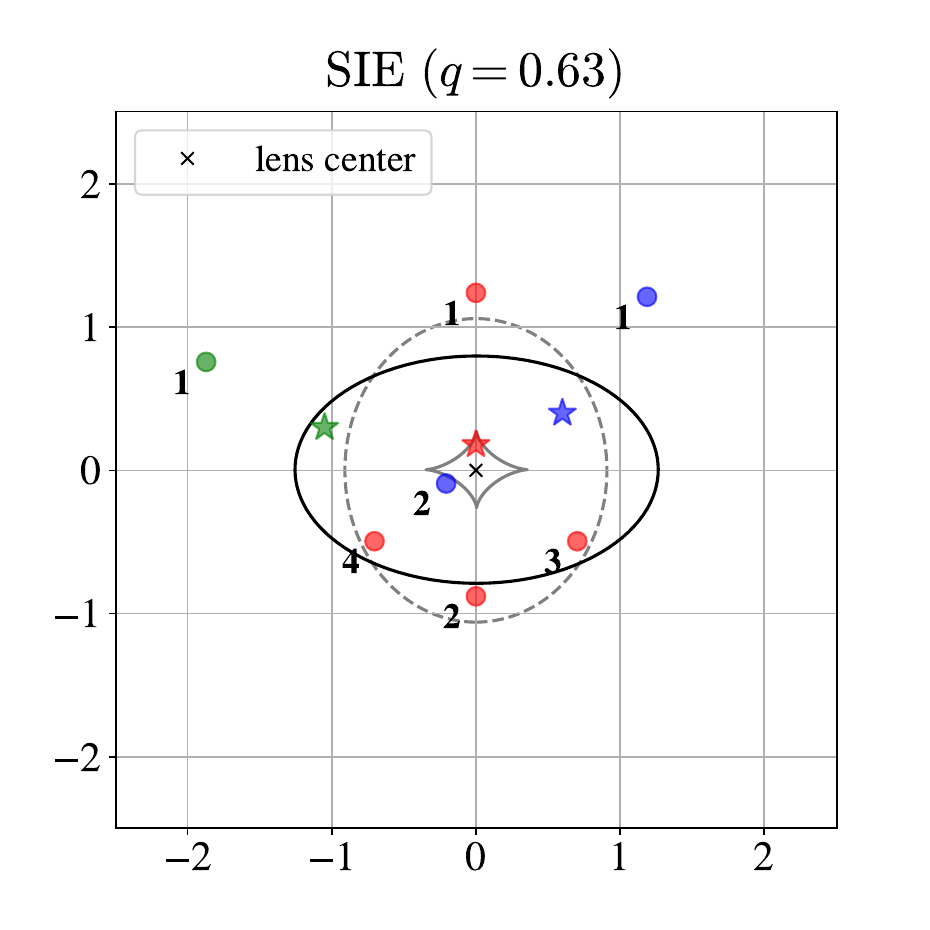}
    \includegraphics[width=0.33\linewidth]{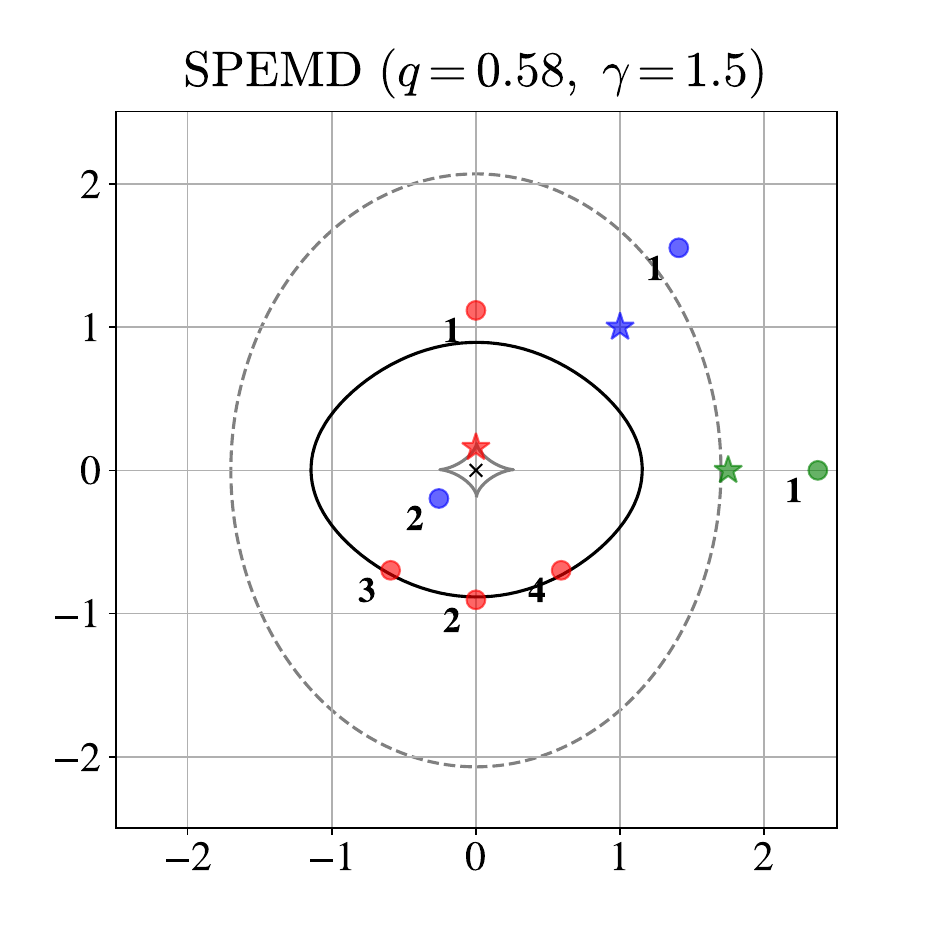}
    \\
    \includegraphics[width=0.33\linewidth]{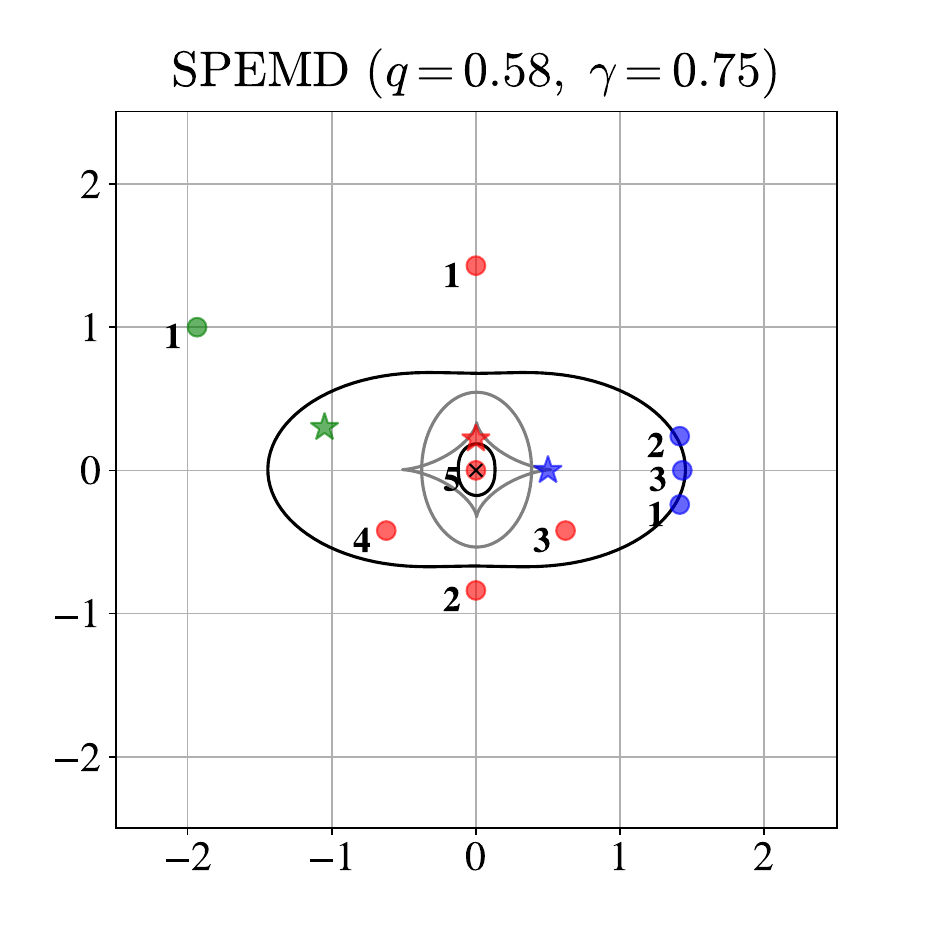}
    \includegraphics[width=0.33\linewidth]{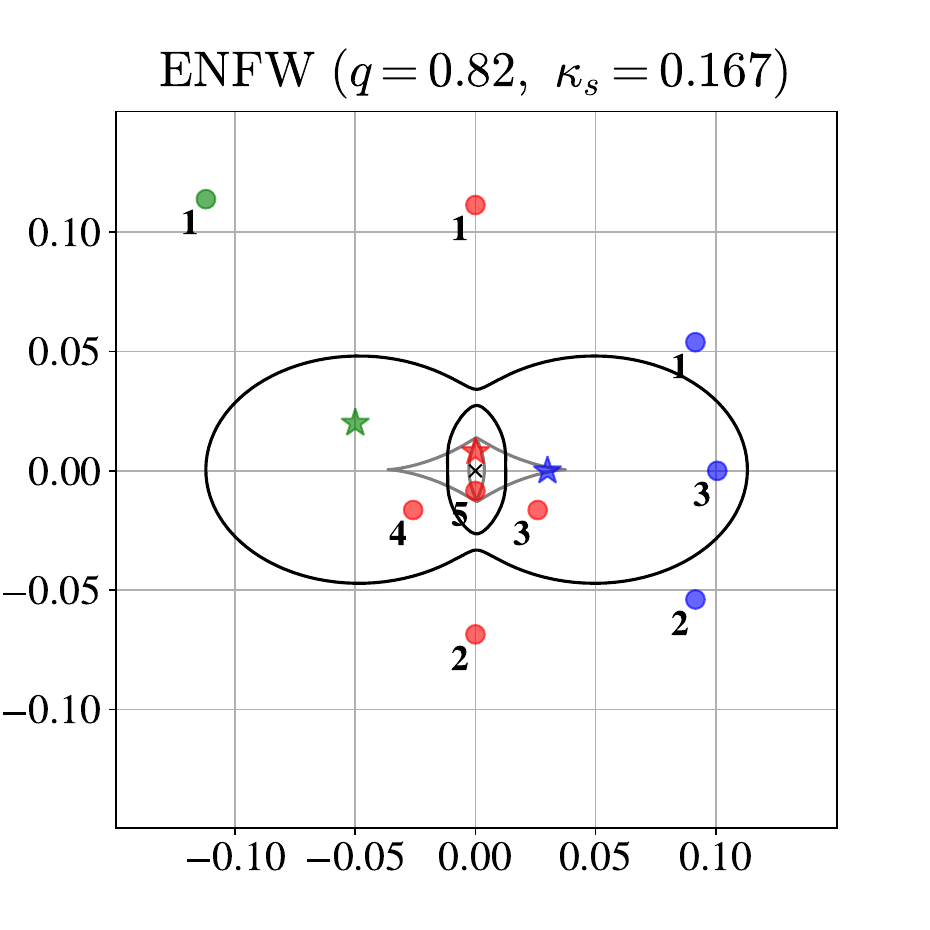}
    \caption{Critical lines (solid black lines), cut (dashed grey lines) and caustics (solid grey lines) of SIE, SPEMD ($\gamma \geq 1$), SPEMD ($\gamma < 1$) and ENFW lenses specified in Table~\ref{tab:inj_lenparams}.
    The same configuration is used as in Fig.~\ref{simple_caustic}.
    Since ellipticity breaks the axial symmetry of the lenses, the point-like tangential caustics turn into a finite area with four cusps.
    The SIE and SPEMD ($\gamma \geq 1$) lenses have a cut because of their singularities at the lens centre.
    When a source is located within the cut, two images are produced, and the two images split into four images when the source falls within the caustic as well.
    On the contrary, the SPEMD ($\gamma < 1$) and ENFW lenses have two caustics.
    If a source lies within one or two caustics, three or five images are created respectively.}
    \label{complex_caustic}
\end{figure*}

Analogous to the SPEMD-like lens model, the normalisation constant for the ENFW model is the same as that of the NFW model.
Also, the point-like tangential caustic transforms into a curved region with cusps due to ellipticity. 
Depending on whether or not the source is located within caustics, single, triple or quintuple images can be generated.
Fig.~\ref{complex_caustic} shows an example of how a source splits into multiple images by an ENFW lens.

\bibliographystyle{aasjournal}
\bibliography{main}
\end{document}